\documentclass[sigconf]{acmart} % camera-ready

\AtBeginDocument{%
  \providecommand\BibTeX{{%
    \normalfont B\kern-0.5em{\scshape i\kern-0.25em b}\kern-0.8em\TeX}}}

\setcopyright{acmlicensed}
\copyrightyear{2018}
\acmYear{2018}
\acmDOI{XXXXXXX.XXXXXXX}

\acmConference[Conference acronym 'XX]{Make sure to enter the correct
  conference title from your rights confirmation emai}{June 03--05,
  2018}{Woodstock, NY}
\newcommand{\etal}{\textit{et al}.}
\newcommand{\nickname}{avaTTAR}
\newcommand{\cmmnt}[1]{}
\newcommand{\onbody}{\textit{\textbf{on-body}}}
\newcommand{\detached}{\textit{\textbf{detached}}}
\newcommand{\Onbody}{\textit{\textbf{On-body}}}
\newcommand{\Detached}{\textit{\textbf{Detached}}}

\usepackage[linesnumbered,ruled,vlined]{algorithm2e}
\usepackage{amsmath}
\usepackage{graphicx}
\usepackage{subcaption}
\usepackage{csquotes}

\usepackage{soulutf8}

\providecolor{modified}{RGB}{65,105,225}   % blue
\providecolor{deleted}{rgb}{1,0,0}     % red
\providecolor{added}{RGB}{65,105,225}   % blue
\newcommand{\added}[1]{{\ifhighlight {{\color{added}{}#1}} \else {#1}\fi}}
\newcommand{\deleted}[1]{{\ifhighlight {{\color{deleted}\st{#1}}} \else {}\fi}}
\newcommand{\modified}[1]{{\ifhighlight {{\color{modified}{}#1}} \else {#1}\fi}}

\newif\ifhighlight
\begin{document}

% \title[avaTTAR]{avaTTAR: Table Tennis Stroke Training with Embodied and Detached Visualization in Augmented Reality}
\title[avaTTAR]{avaTTAR: Table Tennis Stroke Training with On-body and Detached Visualization in Augmented Reality}

\author{Dizhi Ma}
\authornote{Both authors contributed equally to this research.}
\affiliation{%
  \institution{Elmore Family School of Electrical and Computer Engineering \\ Purdue University}
  \city{West Lafayette}
  \country{USA}}
\email{ma742@purdue.edu}

\author{Xiyun Hu}
\authornotemark[1]
\affiliation{%
  \institution{School of Mechanical Engineering \\ Purdue University}
  \city{West Lafayette}
  \country{USA}}
\email{hu690@purdue.edu}

\author{Jingyu Shi}
\affiliation{%
  \institution{Elmore Family School of Electrical and Computer Engineering \\ Purdue University}
  \city{West Lafayette}
  \country{USA}}
\email{shi537@purdue.edu}

\author{Mayank Patel}
\affiliation{%
  \institution{School of Mechanical Engineering \\ Purdue University}
  \city{West Lafayette}
  \country{USA}}
\email{pate1421@purdue.edu}

\author{Rahul Jain}
\affiliation{%
  \institution{Elmore Family School of Electrical and Computer Engineering \\ Purdue University}
  \city{West Lafayette}
  \country{USA}}
\email{jain348@purdue.edu}

\author{Ziyi Liu}
\affiliation{%
  \institution{School of Mechanical Engineering \\ Purdue University}
  \city{West Lafayette}
  \country{USA}}
\email{liu1362@purdue.edu}

\author{Zhengzhe Zhu}
\affiliation{%
  \institution{Elmore Family School of Electrical and Computer Engineering \\ Purdue University}
  \city{West Lafayette}
  \country{USA}}
\email{zhu714@purdue.edu}

\author{Karthik Ramani}
\affiliation{%
  \institution{School of Mechanical Engineering \\ Purdue University}
  \city{West Lafayette}
  \country{USA}}
\email{ramani@purdue.edu}

\renewcommand{\shortauthors}{Ma and Hu, et al.}

\begin{abstract}
Table tennis stroke training is a critical aspect of player development. 
We designed a new augmented reality (AR) system \nickname~for table tennis stroke training. 
The system provides both ``\onbody'' (first-person view) and ``\detached'' (third-person view) visual cues, enabling users to visualize target strokes and correct their attempts effectively with this dual perspectives setup. 
By employing a combination of pose estimation algorithms and IMU sensors, \nickname~
% overcomes practical challenges, 
captures and reconstructs the 3D body pose and paddle orientation of users during practice, allowing real-time comparison with expert strokes. 
Through a user study, we affirm \nickname~'s capacity to amplify player experience and training results. 
\end{abstract}



\begin{teaserfigure}
  \includegraphics[width=\textwidth]{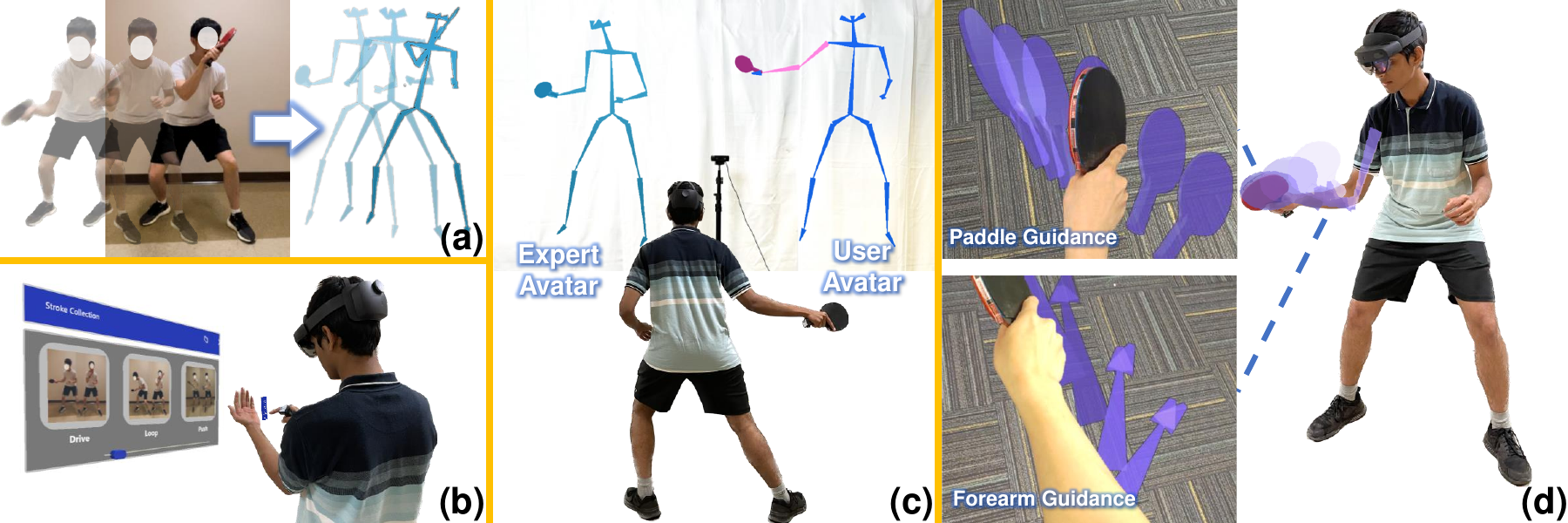}
  \caption{
  Overview of \textit{\nickname}:  An augmented reality system designed to assist table tennis stroke skills learning through \onbody~ and \detached~ visual cues. The expert records their body and paddle movement strokes (a) and saves them in our system. The user selects a specific table tennis stroke (b) during practice, the system provides a separate display of an Expert Avatar (in green) demonstrating the selected stroke alongside the User Avatar (in blue), accurately reproducing the user's real-time posture, and highlighting the error joints (in pink), (c) concurrently, it closely assesses the user's movements, offering on-body guidance (in purple) for (d) optimizing both body and paddle positioning. 
  % An example of our table tennis stroke learning system. Our system provides demonstration and feedback visuals. Demonstration cues: (a-1) detached expert avatar shows target stroke (left) and detached user avatar (right) shows implementation movement; (b-2) embodied expert avatar overlay on user body shows target stroke. Feedback cues examples for arm and paddle correction: (a-2) detached user avatar highlights incorrect skeleton (in pink) and paddle; (b-2) embodied avatar shows arm trajectory; (b-3) paddle trajectory; or (b-4) both.
}
  \label{fig:teaser}
\end{teaserfigure}




\maketitle

\section{Introduction}
% Among the wide range of table tennis techniques, stroke training can be.....
% Table tennis techniques encompass a wide range of elements
Stroke training 
% (the action in which a player strikes a ball) 
can be considered the most fundamental for all players among the wide range of table tennis techniques~\cite{lodziak_2020_spin,artyomutochkin_2018_modern,dubina_2015_100,hodges_2014_table}), as it forms the basis for executing various shots effectively. 
Traditional table tennis stroke training involves coaches who demonstrate specific techniques, players who imitate the strokes, and receive feedback and corrections from the coach~\cite{raab2005improving, wu2010coach, martinent2020literature}. 
Alternatively, players often try to learn strokes by watching online video clips and mimicking the techniques demonstrated. Some players record their own footage and review it for self-improvement. 
Recently, advancements in commercial Virtual Reality (VR) and Augmented Reality (AR) equipment have opened up new opportunities to practice table tennis skills in immersive AR/VR games \cite{vr_2023_eleven, karatas2023consider}. Regardless of which methods to learn strokes, players often struggle with perceiving the correct motion trajectories spatially, either from an expert's demonstration or their own attempts, leading to a vague understanding of the correct movements and how to rectify errors.
% However, coaching methods may face challenges such as trainee self-awareness and skill acquisition through observation --- the ability to see their own performance and translate the coach's execution from a third-person perspective.
% These limitations underscore the need for new training methods that address these challenges and provide intuitive learning opportunities for players.

% user不知道正确的动作轨迹，不知道自己的动作轨迹 -》不知道对错，难以理解动作

In an attempt to help the user understand their motion and feedback on errors, prior research has explored techniques for analyzing video data and displaying the visualization on screen. Researchers have compared user performance with that of experts and provided corresponding feedback using video clips for general motor learning \cite{fieraru2021aifit, chen2018computer, guo2022dancevis, clarke2020reactive} and sports \cite{liu2022posecoach, wang2019ai} with computer vision algorithms. 
Such methods also empowered AR and VR-based sports training systems, which provide more immersive and real-time visual experiences that traditional methods cannot replicate.
Noteworthy examples span a spectrum of sports and physical practices, including basketball~\cite{lin2021towards}, dancing~\cite{anderson2013youmove}, and beyond~\cite{kajastila2016augmented, nozawa2019vr, ye2020shuttlespace}. 
These works typically present visual cues in a third-person view, ``detached'' from the user. 
However, mirroring skills through detached cues may not always be intuitive compared to first-person view ~\cite{yu2020perspective}.
Researchers have addressed this challenge from a first-person perspective in VR \cite{hoang2016onebody}, aiming to increase learner posture accuracy by rendering virtual avatars of both the learner and the expert together.

To this end, we introduce \nickname, an AR system designed for table tennis stroke movement learning.  
Inspired by our preliminary interview with 11 table tennis players, \nickname~ aims to provide a dual-view perspective visualization mechanism by combining on-body visualizations with detached ones, facilitating a better understanding of both the expert's and the user's execution. 
% The system offers user movement feedback for the body and paddle during table tennis stroke training.
The ``\onbody'' visuals overlay virtual cues directly onto the user's physical body, enabling first-person view guidance in AR. 
The "\detached" visuals separate cues from the user, facilitating third-person observation of movements. 
% We offer a dual-view perspective visualization mechanism by combining these \onbody~ cues with \detached~ ones, facilitating a better understanding of both the expert's and the user's execution.
By showing the user and expert execution with avatars and providing \onbody~ and \detached~ visualization, \nickname~ could be used in common table tennis training sessions such as shadow practice~\cite{babar2021analysis, flores2010effectiveness} and multi-ball practice~\cite{gu2019effects}. 

Our system can capture the motion of both experts and users by employing a 3D pose estimation algorithm \cite{mehta2017vnect} and an Inertial Measurement Unit (IMU) sensor to track the paddle orientation.
% We then reconstruct the stroke execution for both experts and users. 
The system provides recording software to record experts' strokes using this visual-sensor solution, storing them in a database.
Meanwhile, the system also provides an AR sub-system for training, when using the AR system for training, users can then choose specific strokes to practice, with two detached avatars displayed during training: one representing the expert's recorded demonstration and the other reflecting the user's real-time execution. Users could change the viewpoint of the detached avatars in whatever way they prefer. 
% Additionally, the user could overlay an on-body expert avatar on their body to follow with.
The system continuously compares the user's motion with the expert's, offering detached feedback that highlights incorrect joints or the paddle and corresponding on-body trajectory guidance to help correct their strokes. 
% We conducted a user study showing that \nickname's learning outcome improved compared with the video approach. 
% We further discussed the hardware and software challenges of our system, deriving insights for future works.
% ver 2
% Our system has an authoring mode to record the expert's execution and a training mode to demonstrate the stroke and provide feedback for stroke training. The recording mode is designed to be
% easy to use, so anyone can record movement sequences and edit them for learning, without the need for complicated motion capture hardware or software. The training mode generates 3D reconstruction of both the expert and the user, providing detached and embodied visualization. 
% The detached cues are adjustable third-person view stroke demonstration and user execution with feedback of incorrect parts, while the embodied feedback provides overlaid visual cues on the user's body for stroke demonstration and guidance for the incorrect parts.
% To empower the 3D reconstruction, we employ a visual-sensor solution to capture the 3D body pose and paddle orientation.

In summary, we contribute:
\begin{itemize}
    % \item The design rationale for table tennis stroke training derived from formative interviews with experienced table tennis players.
    % \item A table tennis stroke training system that provides demonstration and feedback enables body-paddle movement in Augmented Reality.
    % \item A dual view perspective mechanism with embodied and detached visualization that allows comprehensive pose and paddle movement learning.
    % \item A table tennis learning system that provides stroke demonstration, and feedback for self practice with embodied and detached visualization. 
    % To effectively capture body and paddle 3D movements for empowering aforementioned functions in real-time, we also introduced a visual-sensor solution. 
    % \item An AR interface is implemented to enable users to interact with visual cues in a personalized way, enhancing the overall training experience.
    % \item A user study demonstrates the educational efficacy of using detached demonstration and embodied guidance, and another user study compares our system with traditional video-based learning.
    \item A design rationale for table tennis stroke training, informed by formative interviews with experienced table tennis players.
    \item A table tennis stroke training system in AR, incorporates a dual-view perspective visualization mechanism, featuring on-body and detached visual cues, enabling an intuitive experience of table tennis skills acquisition.
    \item A user study showcasing the educational effectiveness of the system over a traditional video-based learning method, together with a separate study evaluating the usability of the system.
\end{itemize}

\section{Related Work}

\subsection{Sports Data Visualization}
With the advent of online video-sharing platforms, social media, and portable devices equipped with cameras, motion analysis methods for sports have become increasingly prominent due to their accessibility and convenience. In team sports, the tactical dimension has also been addressed through visual analytics systems that unveil winning strategies~\cite{stein2017bring, stein2018revealing}.
The evolution of deep convolution networks and algorithms in computer vision has dramatically expanded the potential for analyzing sports data within videos and delivering feedback, encompassing the tracking of objects like balls and subjects like players. Video-based visual analytics systems have been used to enhance game videos~\cite{chen2021augmenting} and to annotate videos~\cite{deng2021eventanchor}.
These techniques also facilitate posture comparison among different players. Guo \etal~presented DanceVis~\cite{guo2022dancevis}, which employs a deep learning pose estimation model to analyze videos, thus enhancing analysis efficiency and automatically yielding individual global performance curves.
Similarly, Wang \etal~\cite{wang2019ai} developed a coaching system for skiing. This system tracks pose trajectories and identifies anomalous poses to provide corresponding "good examples."

These prior works primarily rely on 2D pose estimation. While 2D pose estimation is effective in many cases, it may lead to limitations~\cite{wang2021deep, sarafianos20163d, liu2022recent} in accurately capturing the full-depth and three-dimensional aspects of complex movements.
By generating a comprehensive dataset featuring 3D human poses for fitness training, Fieraru \etal~\cite{fieraru2021aifit} introduced AIFit. This system leverages the dataset for training purposes, offering valuable posture comparison natural language feedback base on the 3D body pose analysis.
Liu \etal~\cite{liu2022posecoach} introduced PoseCoach, which offers customized visualization and feedback based on the specified pose attributes from running videos. PoseCoach also relies on 3D pose analysis. However, this customization lacks real-time interaction and immediate feedback, restricting skill refinement.
Both AIFit and PoseCoach utilize vision-based methods to acquire 3D information. Motion sensing devices could also be employed to capture 3D body poses. Clarke's ReactiveVideo~\cite{clarke2020reactive} aligns experts' poses from a Microsoft Kinect device with novices' poses in videos, adjusting playback based on user proficiency. 

Recognizing existing gaps, such as the absence of prompt feedback in current methods and the increasing inclination towards 3D pose analysis, we are motivated to build an AR system that presents a more intuitive and effective solution for sports motor skill training, especially in the context of table tennis.

\subsection{XR-based Sports Training}
Using XR for sports training taps into the power of interactive visuals and simulated environments, offering athletes opportunities to learn, practice, and refine their skills. In non-traditional sports, XR-based training systems have shown incredible potential. Nozawa \etal~\cite{nozawa2019vr} innovated an indoor VR ski training system that showcases the movements of professional players while facilitating comparisons between users and these seasoned athletes. Kajastila \etal~\cite{kajastila2016augmented} devised an augmented climbing wall, incorporating it with three interactive applications. Ikeda \etal~\cite{ikeda2019golf} project users' postures as virtual shadows on the ground, providing the corresponding feedback.

In the area of general physical training and conventional sports, Han \etal~\cite{han2017my} developed a VR system for self-practicing Tai Chi Chuan, employing an optical see-through head-mounted display (HMD). This setup allows users to be surrounded by multiple virtual coaches, with the added flexibility to adjust perspectives. Prior research suggests that while demonstrating the coach's movements is essential for skill transfer to the trainee, displaying the trainee's own movements is equally crucial in enhancing their self-awareness.
Chan \etal~\cite{chan2010virtual} utilized a wearable sensor-based motion capture suit to document users' movements and subsequently rendered them within a virtual environment. Analogously, Takahashi \etal~\cite{takahashi2019vr} employed wearable sensors to capture and reconstruct three-dimensional body poses, then trained baseball batters. However, these wearable sensor approaches, while potentially yielding precise estimations, inevitably introduce inconveniences and complexities since the wearable suits and sensors can be cumbersome or uncomfortable to wear, potentially affecting the natural movement of the user.
Additionally, RGBD sensors, such as Microsoft Kinect, were used by Anderson \etal~\cite{anderson2013youmove} in their YouMove system. This system empowers experts to record sequences of physical movements, and novices to practice and learn them using an AR mirror. 
% It is particularly noteworthy that its demonstrated efficacy in enhancing movement learning experience and outcomes, is evidenced within the context of ballet dance evaluations. 
In a similar process, Zhou \etal~\cite{zhou2022movement} delved into the effectiveness of movement acquisition through a mixed reality mirror.

These prior studies displayed feedback adjacent to or encircling the user, adopting a third-person perspective. While this perspective can offer valuable insights, a first-person viewpoint (FPV) has the potential for more intuitive visualization \cite{yu2020perspective, lin2021towards, han2016ar, hoang2016onebody}. 
In recent research on basketball, an in-situ display approach~\cite{lin2021towards} was explored, in which basketball trajectories were visualized through an AR head-mounted display, offering trajectory visualization directly within the user's field of view.
For motor skill learning, Han \etal~\cite{han2016ar} investigated FPV for arm movement learning, and \cite{yu2020perspective, hoang2016onebody} utilized FPV for movement learning in Virtual Reality (VR), enabling users to observe both their own virtual skeleton and the teacher's virtual skeleton.  
In this paper, we introduce an embodied approach in AR, in which the virtual body is superimposed on the physical one, allowing the user to actively conform to the template.

% Building upon the foundations laid by the prior studies, we seek to introduce a new approach. 
Our objective is to offer users a comprehensive visualization experience by juxtaposing both \detached~ expert and user avatars while incorporating \onbody~ cues for movement correction. 

\subsection{Training Systems for Table Tennis}
In table tennis, computer-aided methods have advanced to make gameplay analysis easier. These methods include an annotation tool~\cite{deng2021eventanchor} that integrates with computer vision algorithms, thus enhancing the analysis of the dynamics of table tennis. Additionally, a data visualization tool~\cite{chen2021augmenting} has emerged to provide insight into game video analysis from the comfort of one's home. Another distinct example is Wang's work~\cite{wang2022tac}, which adapted the Internet of Things (IoT) devices to track arm and paddle movements. The resulting training system interprets technical attributes and the corresponding indicator values, presenting these data through an intuitive software interface.

In VR applications, earlier work like that of Brunnett \etal~\cite{brunnett2006v} built immersive table tennis simulation systems. However, their focus leaned towards constructing realistic virtual environments rather than optimizing study efficiency. Subsequent research~\cite{michalski2019getting, wu2021spinpong, oagaz2021performance} has shed light on the effectiveness of training in VR landscapes. Michalski \etal~\cite{michalski2019getting} conducted tests with a VR game, while Wu \etal~\cite{wu2021spinpong} focused on providing multimodal cues to improve the successful return rate. Oagaz \etal~\cite{oagaz2021performance} indicated players' incorrect postures by highlighting the joints of the user's reconstructed skeleton in the virtual environment. These collective efforts paint a multifaceted picture of training possibilities within VR-based table tennis interventions.
However, current VR simulations suffer limitations \cite{karatas2023consider} including the absence of genuine physical and precise movement feedback, even in the latest table tennis VR games~\cite{vr_2023_eleven}.
Shifting the focus to augmented reality initiatives, the origins trace back to~\cite{ishii1999pingpongplus}, which introduced a reactive AR table. Subsequent work, including those by Mayer~\cite{Mayer_2019}, also embraces AR tables, offering enhanced visualization.  
These advances have excelled in effectively showing the results of various techniques, including ball trajectory, ball placement, and return rate. However, it is important to highlight that these often overlooked the crucial elements of motor skill learning, specifically focusing on the player's body and paddle, and lacked essential guidance.

In this work, our main objective is to assist the stroke training of table tennis by integrating motion guidance through virtual avatars in AR. 
% Leveraging augmented reality techniques, we aim to bridge the gap between visualization and practical execution, ultimately enhancing the learning experience for both body and paddle motion while providing intuitive real-time guidance.

\section{Design Rationale}\label{section:DesignRationale}
\begin{figure*}
  \centering
  \includegraphics[width=\linewidth]{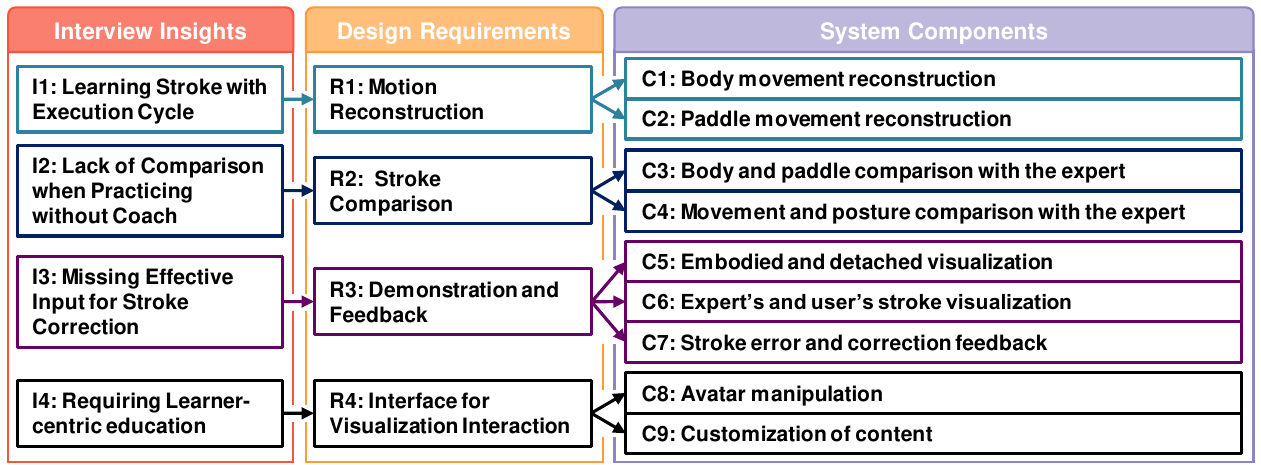}
  \caption{Illustration of design requirements and system components. We have extracted four key design requirements from our formative interviews with individuals experienced in table tennis training. In response, we have elaborated on these requirements, resulting in the development of nine detailed modules for implementation.}
  \label{fig:Figure_DesignRationale}
\end{figure*}

We conducted 
\deleted{a survey}
\modified{interviews}
with expert players and table tennis coaches (Section \ref{FormativeInterview}), to understand their training experience, challenges and pain points, training methods, and coaching experience (if any). We use these data and the results of these interviews to derive the design requirements for our system (Section \ref{DesignRequirements}). A summary of the insights, requirements, and subsequent components is shown in ~\autoref{fig:Figure_DesignRationale}.

 \subsection{Formative Interview}\label{FormativeInterview}
Our methodology to obtain the insights began with formative interviews involving a diverse cohort of 11 table tennis players. 
These interviews were divided into several sections, each focusing on critical aspects of their table tennis experience. 
% The insights extracted from these interviews offer a discipline into the landscape of table tennis training.

\added{
We conducted in-person interviews with each interviewee individually. 
The interviews were semi-structured with open-ended questions. 
The entire interview procedure lasted between 15 to 30 minutes, depending on the interviewees. 
We recorded the audio of all conversations and converted them into text for analysis.
}

The demographics of the interviewees are as follows: the average age was 38.6 years (SD = 19.2), and their cumulative experience in table tennis averaged 20.0 years (SD = 14.7). Of the participants, seven had prior experience practicing under the guidance of a coach, with an average coaching duration of 4.7 years (SD = 4.5). Only two participants were familiar with AR, and none had used AR for table tennis practice.

\subsubsection{Interview Topics}

During the formative interviews, we explored several key topics, including:

\begin{itemize}
  \item Table Tennis Training Experience. Participants were asked to share their personal experiences with table tennis stroke training, including any formal coaching or self-guided practice.
  
  \item Challenges and Pain Points. We inquired about the specific challenges and difficulties they encountered during stroke training, seeking to identify common pain points.
  
  % \item Training Methods. Participants were prompted to describe their preferred methods and resources for learning and improving table tennis techniques.
  
  \item Coaching and Mentorship. For those with coaching or mentoring experience, we delved into their coaching methodologies and the impact of coaching on trainees' performance.
  
  % \item Use of Technology. We explored whether participants had previously used technology or applications to aid in their table tennis training and their perceptions of technology's role in this context.
\end{itemize}

\subsubsection{Findings and Insights}

In our conversations with the interviewees, we gained valuable insights. These insights helped us understand the details of stroke training and the difficulties that players encounter when training alone. 

\textbf{I1: Learning Stroke with Execution Cycle.} One of the fundamental insights that emerged from our interviews relates to the essence of stroke training in table tennis. Most strokes can be deconstructed into three distinct stages: the ``prepare - back swing - fore swing - recovery'' cycle~\cite{artyomutochkin_2018_modern,dubina_2015_100}. The key to executing a correct stroke lies in mastering the precise positions and transitions within these stages and translating them into a smooth motion. Our interviewees emphasized that the path to improvement involves rigorous repetition to form muscle memory.

\textbf{I2: Lack of Comparison when Practicing without Coach.} A significant portion of table tennis training occurs without the presence of a coach. Although self-practicing is essential due to the sheer volume of repetition required to master the sport, it comes with its own set of challenges. Specifically, without a coach's guidance, it becomes difficult to discern when a stroke is being executed incorrectly. Even more concerning, continued practice without correction can lead to the formation of muscle memory for incorrect motions, making subsequent corrections more challenging.

\textbf{I3: Missing Effective Input for Stroke Correction.} Coaching and mentorship were highlighted as invaluable resources for stroke training. Coaches can provide multifaceted assistance, including:
% \begin{itemize}
%     \item {\textbf{\textit{Demonstration}}:} Coaches can effectively demonstrate the correct movement and stroke execution, offering a visual reference for players.
%     \item {\textbf{\textit{Observation}}:} Coaches naturally know how to closely watch and break down the small details of a player's movements.
%     \item {\textbf{\textit{Feedback}}:} Coaches have the unique capacity to provide immediate, tailored, and constructive feedback during practice sessions. 
% \end{itemize}
{\textbf{\textit{Demonstration}}:} Coaches can effectively demonstrate the correct movement and stroke execution, offering a visual reference for players.
% 2) {\textbf{\textit{Observation}}:} Coaches naturally know how to closely watch and break down the small details of a player's movements.
{\textbf{\textit{Feedback}}:} Coaches have the unique capacity to provide immediate, tailored, and constructive feedback during practice sessions. 
However, training with a coach may still present challenges in transferring skills, as bridging the gap from observing third-person movements to adopting an egocentric perspective can be complex.

\textbf{I4: Requiring Learner-centric education.} During training, individuals have diverse preferences for the content they wish to observe. Different strokes may require distinct viewing angles, and the specific focus of observation varies from person to person. For example, some may prioritize analyzing arm and footwork, while others may concentrate on monitoring paddle positioning and rotation.

In the following section, we outline the specific features and functions of our system that directly address the identified needs and challenges voiced by the interview participants.

\subsection{Design Requirements}\label{DesignRequirements}

From the initial interviews with the potential user with table tennis training and coach experience, as well as the limitations of existing works, we identify the following design requirements:

\subsubsection{R1: Enable body-paddle motion reconstruction.}
% body movement reconstruction.  
% paddle movement reconstruction.  
% To provide a comprehensive training analysis, our system developed the motion reconstruction module. 
% Building on \textbf{I1}, which emphasizes the importance of mastering the precise stages of stroke execution, our first design requirement aims to facilitate stroke training by enabling motion reconstruction.
% This module comprises two crucial components:

Based on the insight from \textbf{I1}, which emphasizes the significance of mastering precise stroke stages, our first design requirement, is aimed at facilitating stroke training. It comprises two essential components:

\begin{itemize}
  \item C1: Reconstruction of body movements to assess posture and positioning during strokes. According to the interviewee's feedback: \textit{``... I would like to see the movement for the \textbf{body and the paddle} ...''}
  \item C2: Thorough reconstruction of paddle movements to scrutinize the subtleties of stroke execution.
\end{itemize}

\added{We followed YouMove~\cite{anderson2013youmove} and used the skeleton for explicit visualization of the body since YouMove also includes full-body movement learning.}

% These fundamental features will allow the system to obtain and compare users and coaches to gain profound insights into the finer aspects of their performance and enable the remaining modules.

\subsubsection{R2: Integrate stroke comparison.}
% body and paddle comparison with the expert.
% movement and posture comparison.
% In the pursuit of effective stroke comparison within our system for table tennis,
% \textbf{I2} highlights the significance of effective input from a coach for stroke correction. To bridge the gap between coaching and practice, our second design requirement introduces stroke comparison as a core feature, 
% We've identified two core components:
Derived from insights \textbf{I2}, which highlights the challenges of practicing without a coach, our second design requirement, introduces a core feature to address these challenges. This requirement aligns with the need to bridge the gap between coaching and individual practice. We've identified two core components:

\begin{itemize}
  \item C3: A side-by-side comparison of the user's body and paddle movements with those of an expert, allowing the user to observe the errors for improvement. As expressed by an interviewee: \textit{``...So when you have a coach, I mean, a coach can constantly give you feedback ... someone can \textbf{tell you that you did this wrong right at that moment}''}
  \item C4: Comparative analysis of overall movement and important posture about established standards and best practices, helping users strive for excellence. In the words of an interviewee: \textit{``...like suppose I play a stroke if it tells me exactly how I should correct it and \textbf{what my angle is, what the angle should be of an expert}''}
\end{itemize}
% These components will serve as a valuable tool for players and coaches alike.

\subsubsection{R3: Demonstration and feedback visualization.}
% embodied cues for body and paddle correction.
% detached cues to show incorrect parts. 
% humanoid avatars with paddles for both expert and user.

% In our AR-based Table Tennis Simulator, addressing the need for movement corrections is paramount, encapsulated in the requirement of Cognizance. 
% Addressing \textbf{I3} and \textbf{I4}, which underscores the challenges of practicing without a coach as well as the lack of self-observation, our third design requirement focuses on stroke and feedback visualization. 
% This multifaceted requirement is subdivided into three key elements:
In response to insights gathered from \textbf{I3}, which emphasises the role of coaching and mentorship in stroke training, our third design requirement focuses on addressing the challenges faced when practicing without a coach. This multifaceted requirement is designed to add to awareness during movement execution and encompasses three key elements:

\begin{itemize}
  % \item C5: Real-time embodied cues that guide users in making immediate corrections to their body and paddle positioning during training sessions. According to an interviewee's statement: \textit{``... and I would like to see the coach's movement in \textbf{the first person perspective} as well ...''}
  % \item C6: Detached cues highlighting specific aspects of a motion sequence that require attention or correction, ensuring that no detail goes unnoticed. As articulated by an interviewee: \textit{``... see what you are doing wrong from a third point of view, like a 3D recreation ...''}
  % \item C7: Incorporation of humanoid avatars equipped with paddles for both expert and user perspectives, providing a holistic view of stroke execution. As described by an interviewee: \textit{``... having the corrections for the \textbf{body and the paddle} should be there ...''}
  \item C5: A two-view perspective, presenting information in both \detached~ and \onbody~ formats. This versatile approach allows users to receive demonstrations and feedback in various ways, catering to individual learning preferences. As articulated by an interviewee: \textit{``... see what you are doing wrong from a \textbf{third point of view}, like a 3D recreation...''} and another one: \textit{``... and I would like to see the coach's movement from \textbf{his view} as well...''}.
  \item C6: Visual cues that not only demonstrate the ideal execution of the target stroke, including paddle and full-body movement but also provide a simultaneous representation of the user's execution. 
  \item C7: Visual cues not only to locate areas where the user made mistakes but also to offer clear guidance on how to refine their posture. 
\end{itemize}

% These visualization elements will empower users to identify areas for improvement with greater clarity.

\subsubsection{R4: Interface for visualization interaction.}
% avatar manipulation.
% visualization customization.  
% This requirement underpins our aim to craft a user-centric AR system for table tennis training, allowing the user to customize the training focus and enhance the training experience. 
Incorporating insights from our interviews, particularly \textbf{I4}, which underscores the diverse perspectives individuals have during training, our fourth design requirement aims to address the challenges related to catering to varied visualization. This requirement underlines the need to create an intuitive interface that accommodates different viewing preferences, aligning with our user-centric approach. It comprises two pivotal modules:

\begin{itemize}
  \item C8: Intuitive avatar manipulation, allowing users to interact with the visualized data, providing a sense of control over the analysis process. In the interviewee's own words:
  \textit{``... So like, you can slow down time, \textbf{watch and play the stoke in slow motion} ...''}
  \item C9: Customization options that enable users to tailor the visualization to their specific needs, accommodating different skill levels and training objectives. Based on an interviewee's input: \textit{``... And I would like to \textbf{change the viewpoint of the expert}...''}
\end{itemize}
% A user-centric interface that incorporates these design requirements will make the training and analysis process intuitive, more accessible, and engaging.

\section{\nickname~ System Walk-through}

We now elaborate on an example of using our system.
Initially, an expert will use our authoring system to record various stroke movement videos and derive corresponding motion data. 
The expert could further trim and mark the keyframes for these videos, meanwhile, motion data are changed accordingly. 
The edited video and motion data will be saved in our AR system stroke database.
When using our AR system, the system will first calibrate according to the user's height to adjust for the scale of the \onbody~ and \detached~ visualization cues. 
The user can then choose a desired stroke from the database to practice (as depicted in ~\autoref{fig:teaser}(b)).
Subsequently, the system proceeds to visually depict the user's movements through a User Avatar (designated as UA). 
The UA accurately reflects the real-time posture, encompassing the entire body and the paddle's orientation (R1). 
Following this stage, the user selects the specific stroke they wish to learn, prompting the Expert's Avatar (referred to as EA) to appear correspondingly (as illustrated in ~\autoref{fig:teaser}(c)).
Additionally, the user closely observes and emulates the movements of EA, thus acquiring proficiency in executing the stroke. Throughout the learning process, the system compares the user's movements and the ideal model represented by EA (R2). Any deviations or errors in the user's actions are highlighted on UA and the corresponding \onbody~ guidance appears.
This guidance is provided through visual representations, which include movement trajectories highlighting errors in body parts and paddle positioning. These are superimposed on the user's body based on the parameter of UA, aligning with the initially established reference points (R3, as shown in ~\autoref{fig:teaser} (d)). 
During training, users have the flexibility to manipulate the position, viewpoint, and scale of both UA and EA. Furthermore, they can adjust the playback speed of EA, allowing for a ``slow-mo'', detailed examination of the stroke. 
All of these interactions are seamlessly facilitated through our user-friendly interface (R4).

\begin{figure*}[h!]
  \centering
  \includegraphics[width=\linewidth]{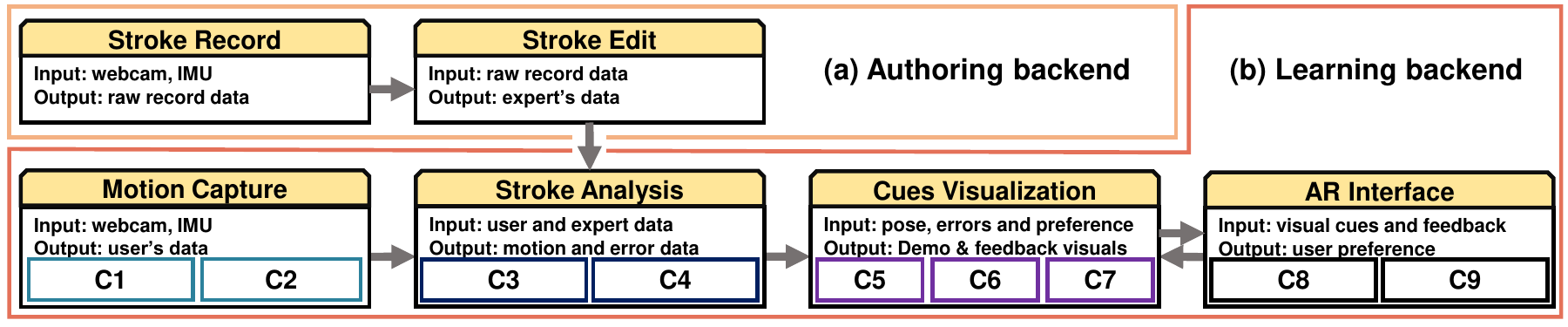}
  \caption{
  % Illustration of the system's back-end workflow. The motion capture module utilizes a webcam and an IMU input to derive the user's body and paddle pose. The stroke analysis module compares user and target movements, generating error scores. Additionally, the cues visualization module employs pose, error data, and visual preferences to provide detached and embodied feedback visualizations. Users could adjust the visualization via the AR interface.
  System's backend workflow. (a) The authoring backend: enables the expert to record and edit their movement posture data. (b) The learning backend: captures the user's motion and compares it with expert data, providing visual cues for stroke learning, and allows the user to adjust these visuals via the AR interface.
  }
  
  \label{fig:SystemWorkflow}
\end{figure*}

\section{\nickname~ Authoring System}
\begin{figure}[h!]
  \centering
  \includegraphics[width=\linewidth]{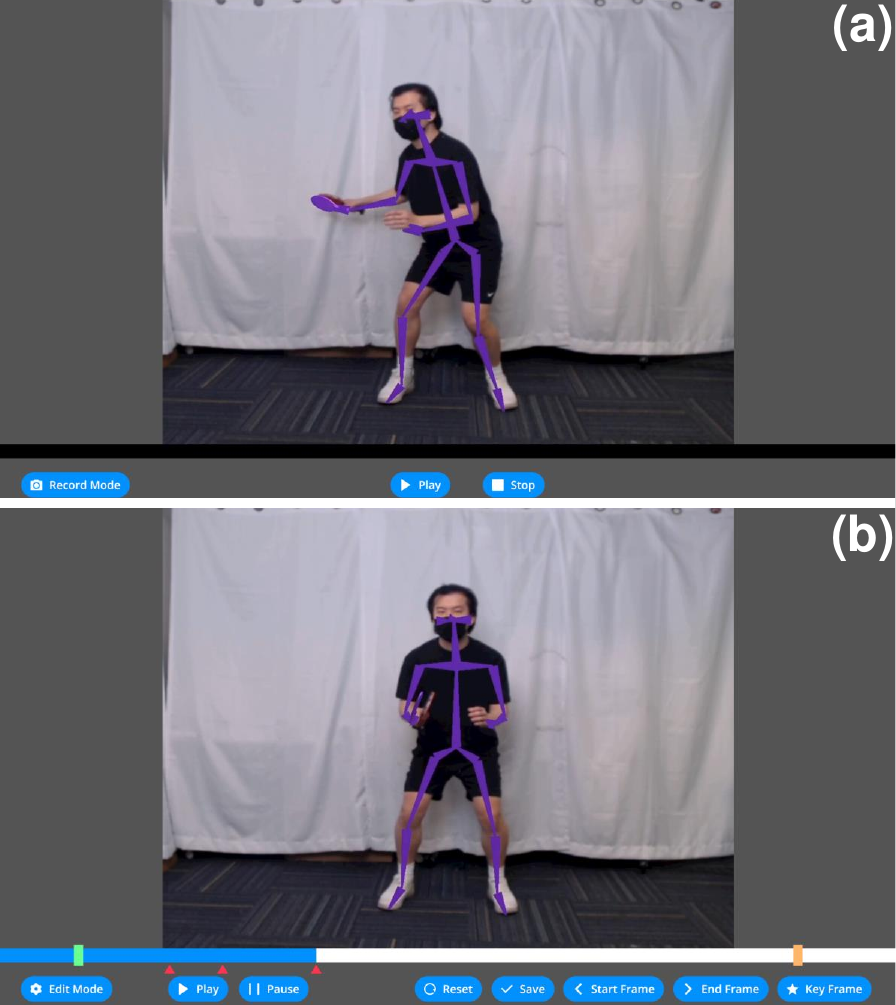}
  \caption{The authoring interface: (a) The record mode allows the user to record video and stroke. (b) The edit mode interface allows authors to trim video and specify keyframes.}
  \label{fig:Aurhor}
\end{figure}

We developed a simple interface similar to \cite{anderson2013youmove} as shown in \autoref{fig:Aurhor} for the expert to capture stroke movement and annotate key posture. 
% To simplify the capture process, \nickname allows experts to contribute learning material without the need for complicated motion capture hardware or software. 
The record mode~(\autoref{fig:Aurhor}(a)) allows authors to record themselves performing the movement. The system captures the 3D body movement and paddle orientation data of the author. 
After starting the software in record mode, the expert is presented with a screen that has a ``Start'' button, a ``Stop'' button, as well as the current video stream from a webcam. To capture movement, the expert presses the ``Start'' button, performs the stroke, and then presses the ``Stop'' button. Note that we attached an IMU to the paddle to capture the orientation of the paddle.
The edit mode~(\autoref{fig:Aurhor}(b)) allows the expert to trim the recording to remove unwanted data with the ``Start Frame'' and ``End Frame'' buttons. 
The authors then specify the keyframes for the recorded movement with the ``Keyframe'' button. For our user study, keyframes are postures that reflect the key stages mentioned in Section \ref{section:DesignRationale}, which helps the user understand the stroke in the big picture. The green rectangle marker on the video process bar reflects the start frame, the brown marker reflects the end frame, and the red triangle markers reflect the keyframes. The user could reset all markers with the ``Reset'' button.

\section{\nickname~ Learning System}
The \nickname~ learning backend comprises several modules as shown in \autoref{fig:SystemWorkflow} (b), these modules are:
Motion Capture, Stroke Analysis, Cues Visualization, and AR Interface.
% \textbf{Motion Capture}: This module captures and tracks the user's movements in real time using a combination of hardware and software.

% \textbf{Stroke Analysis}: In this module, the system analyzes the user's movements and compares them to an expert model, providing insights into the accuracy and alignment of the user's actions.

% \textbf{Cues Visualization}: This module presents the visuals to the user, including detached cues (expert and user avatars) and embodied cues (correct joint movement trajectory and paddle trajectory). These cues involve demonstrating the execution of the expert and the user as well as providing feedback for the user.

% \textbf{AR Interface}: The AR Interface serves as the user's gateway to the system, allowing them to interact with avatars, select specific strokes to learn, and receive feedback and guidance throughout their training.

\subsection{Motion Capture}
% Motion capture is the fundamental module part of our system, results from the motion capture solution allow us to reconstruct the 3D avatar for the player and provide the pose data to analysis. To capture the player's movement, we implemented a solution to estimate the 3D pose of the player and the paddle. Our solution is a real-time markerless 3D motion capture system consisting of the hardware part to obtain the necessary data and software to process the data.
Motion capture constitutes the foundational module of our system. The outcomes derived from our motion capture solution enable us to rebuild a 3D avatar for the player and furnish pose data for analysis. To track the player's movements, we developed an approach to gauge the 3D posture of both the player and the paddle (C1, C2). It comprises hardware for data acquisition and software for data processing.

\subsubsection{Hardware Configuration.} 
\begin{figure}[h!]
  \centering
  \includegraphics[width=\linewidth]{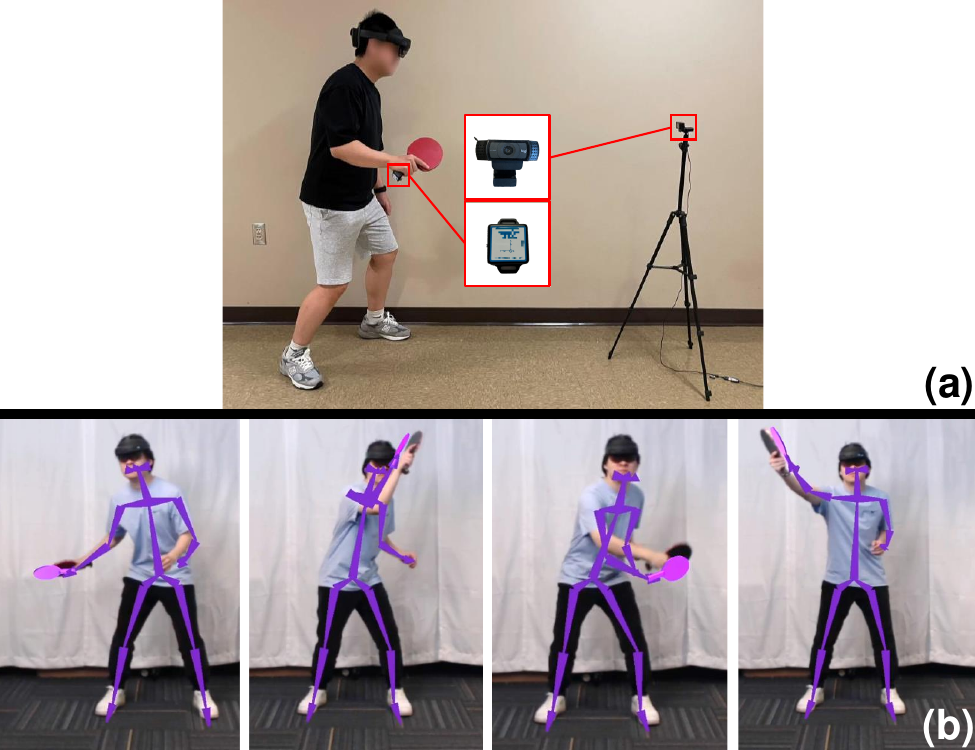}
  \caption{(a) Our motion capture module consists of one webcam placed in front of the user and one IMU attached to the table tennis paddle. The webcam provides the visual input for the system and the IMU provides the paddle orientation. (b) Examples of motion capture results with different body and paddle postures. }
  \label{fig:HardWare}
\end{figure}

Our system relies on a webcam placed in front of the player for optimal data capture, providing an unobstructed view of movements and reactions (\autoref{fig:HardWare}). Additionally, we integrate an IMU on the player's paddle (similar to \cite{wang2022tac}), enabling real-time recording and analysis of paddle pose. 

\subsubsection{Software Setup.} Using a real-time processing 3D pose estimation model, the webcam captures precise 3D joint locations, creating a detailed 3D avatar that mirrors the player's movements. Simultaneously, the IMU on the paddle provides pose data of the paddle.% synchronizing paddle orientation with the player's hand-wrist position.

\subsection{Stroke Analysis}\label{section:stroke_analysis}
 A fundamental requirement is to maintain a consistent posture for users, identical to that of the expert, throughout the stroke learning process. This facilitates a real-time comparison between learner data and expert data extracted from the ongoing frame. In this section, the two algorithms, one for frame-by-frame error comparison (C3) and the other for overall motion error comparison (C4) are presented.
 
\begin{algorithm}[h!]
\caption{Quaternion Dissimilarity-Based DTW for body}
\label{alg:quaternion_dtw_positions}

\textbf{Input:} Joint angle sequences: User's $Q_1$ with sequence length $N$ and expert's $Q_2$ with sequence length $M$, number of joints $J$ \\
\textbf{Output:} DTW distance matrix $D[N][M][J]$

$Q_1 \gets \text{Kalman\_filter}(Q_1)$, $Q_2 \gets \text{Kalman\_filter}(Q_2)$\;
Initialize distance matrix $D$ with dimensions $N \times M \times J$\;

\For{$i = 0$ to $N$}{
    \For{$j = 0$ to $M$}{
        \For{$k = 0$ to $J$}{
            $D[i][j][k] = \infty$\;
        }
    }
}
\For{$i = 0$ to $N$}{
    \For{$j = 0$ to $M$}{
        $D[i][j][0] = 0$\;
    }
}
\For{$i = 1$ to $N$}{
    \For{$j = 1$ to $M$}{
        \For{$k = 1$ to $J$}{
            Calculate quaternion dissimilarity $E = \text{quaternion\_dissimilarity}(Q_1[i][k], Q_2[j][k])$\;
            $D[i][j][k] = E + \min(D[i-1][j][k], D[i][j-1][k], D[i-1][j-1][k])$\;
        }
    }
}
\textbf{Return} $D[N][M][J]$\;
\end{algorithm}

\subsubsection{Body.} 
When studying body movements, we focus on joint angles, calculated from joint positions in the player's body. This data allows us to compare experts and users. We denote the user's joint sequence as ${Q_1}\in \mathbf{R}^{N\times J\times 4}$, where $N$ denotes the number of the sequence and $J$ denotes the number of the joints, and 4 denotes the quaternions rotation. We consider all joints except the end joints, such as the head and toes.

To compare the expert and the user, we further align the user's joint angle series using the Dynamic Time Warping (DTW) algorithm \cite{berndt1994using, muller2007dynamic, muller2007information} with the quaternions dissimilarity equation:
\[ \text{quaternion\_dissimilarity}(q_1, q_2) = 1 - 
% \frac{q_1 \cdot q_2}{\|q_1\| \|q_2\|} 
\langle q_1,~q_2\rangle
\]
where $q_1, q_2$ represent two quaternions vectors. The details of the quaternion dissimilarity-based DTW are shown in Algorithm \ref{alg:quaternion_dtw_positions}. Since the user may act slightly faster or slower than the EA, for real-time comparison, we used $N=M=10$ to reduce the computation cost, otherwise, $N$ and $M$ refer to the length of the sequence. We introduce universal thresholds $\xi_{\text{joint}}$ for all joints to identify incorrect movements.
The user's movement is considered incorrect at joint $k$ if: 
\[ D[N][M][k]/N < 1- \xi_{\text{joint}}. \]

\begin{algorithm}[h!]
\caption{Quaternion Dissimilarity-Based DTW for Paddle}
\label{alg:quaternion_dtw_positions_paddle}

\textbf{Input:} paddle quaternion sequences: User's $Q_1$ with sequence length $N$ and expert's $Q_2$ with sequence length $M$ \\
\textbf{Output:} DTW distance matrix $D[N][M]$

% Convert $P_1$ and $P_2$ to quaternion sequences $Q_1$ and $Q_2$\;%{ using position\_to\_quaternion\};
$Q_1 \gets \text{Kalman\_filter}(Q_1)$, $Q_2 \gets \text{Kalman\_filter}(Q_2)$\;
Initialize distance matrix $D$ with dimensions $N \times M$\;

\For{$i = 0$ to $N$}{
    \For{$j = 0$ to $M$}{
        $D[i][j] = \infty$\;
    }
}
$D[0][0] = 0$
\For{$i = 1$ to $N$}{
    \For{$j = 1$ to $M$}{
        Calculate quaternion dissimilarity $E = \text{quaternion\_dissimilarity}(Q_1[i], Q_2[j])$\;
        $D[i][j] = E + \min(D[i-1][j], D[i][j-1], D[i-1][j-1])$\;
    }
}
\textbf{Return} $D[N][M]$\;
\end{algorithm}

\subsubsection{Paddle.} In contrast to the human body, we can directly use the quaternions received from the IMU to calculate the angle difference using DTW with a quaternion dissimilarity equation. Similarly, a threshold $\xi_{\text{paddle}}$ is implemented to determine feedback intervals. 
The user's paddle movement is considered incorrect if: \[ D[N][M]/N < 1- \xi_{\text{paddle}}. \]

\subsection{Cues Visualization}

\begin{figure}[h!]
  \centering
  \includegraphics[width=\linewidth]{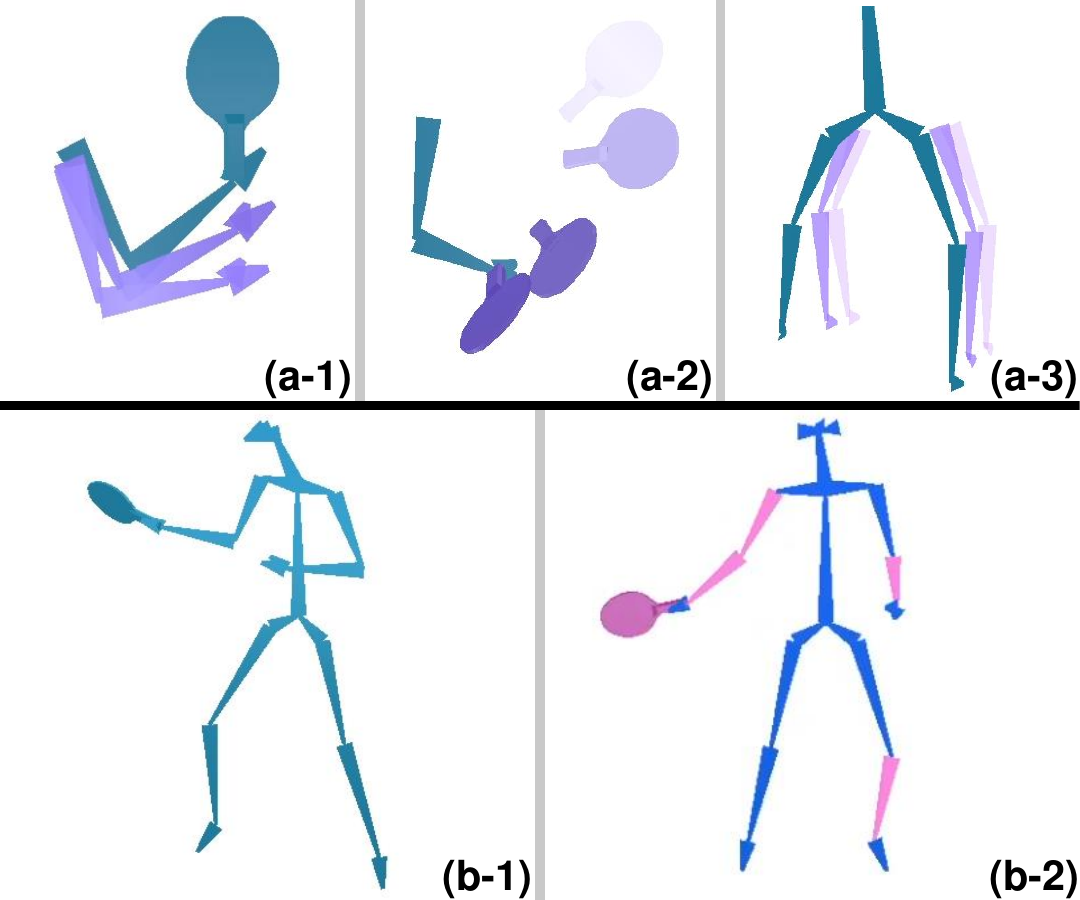}
  \caption{Showcase of our visualization. \Onbody~ cues: (a-1) movement trajectory; (a-2) paddle shadow shows paddle trajectory; (a-3) footwork trajectory. \Detached~ cues: (b-1) expert avatar shows correct movement; (b-2) user avatar shows user's movement and highlights incorrect skeleton (in pink). }
  % \caption{An example for movement guidance visualization. Demonstration cues: (a-1) detached expert avatar shows target stroke (left) and detached user avatar (right) shows implementation movement; (b-1) embodied expert avatar overlay on user body shows target stroke. Feedback cues: (a-2) detached user avatar (right) highlights (in pink) incorrect skeleton and paddle; (b-2) embodied avatar shows arm trajectory; (b-3) paddle trajectory; or (b-4) both.} %(b-5) paddle orientation guide.}
  % \Description{or feedback for different body parts and paddle}
  \label{fig:Visulization}
\end{figure}

As shown in \autoref{fig:Visulization}, we employ two distinct types (C5) of visual cues to enhance the user experience and provide demonstration (C6) and feedback (C7) during execution: \onbody~ cues and \detached~ cues. Both cues are human skeleton-like avatars reconstructed based on the user's and experts' movements of both body and paddle that reflect the difference between them and guidance for the user. The
% \onbody~ cues overlay on the user's physical body, while the \detached~ cues separate from the user.

% JS: what is detached. what is embodied. and why

\subsubsection{Demonstration}
In the demonstration phase of our system, our aim is to visually represent the stroke execution of both the expert and the user during training. The primary goal is not only to showcase the correct movement performed by the expert but also to create awareness for the user regarding their own motion execution.
\begin{itemize}
    \item \textbf{\Detached~ Cues}: For the desired movement, a \detached~ expert avatar illustrates the ideal trajectory. Simultaneously, the user's avatar mirrors their real-time movements, providing a direct comparison. 
    \item \textbf{\Onbody~ Cues}: In addition, the user has the option to overlay an \onbody~ expert avatar onto their physical body. Following the movements of this avatar can help reduce cognitive load, facilitating the transfer of the demonstration into the user's own physical actions.
\end{itemize}

\subsubsection{Feedback}
In the learning phase, the system offers guidance based on the user's execution, employing visualizations to highlight errors and generate instructive cues. The primary objective is to visually indicate areas that require refinement in specific aspects of movement. Similarly to the demonstration phase, the feedback visualization is presented in both \detached~ and \onbody~ formats.
\begin{itemize}
    \item \textbf{\Detached~ Cues}: The backend of the system analyzes the user's movement in comparison to the expert and identifies the error joints on the \detached~ user avatar. These highlighted areas offer clues to where improvements can be made.
    \item \textbf{\Onbody~ Cues}: Concurrently, \onbody~ cues visually represent the correct trajectory of joint movement, providing information on how the user can enhance their execution. This method helps users understand what improvements can be made.
\end{itemize}

\subsection{AR Interface}\label{Section:interface}
% The AR interface has three modes. 
% (1) Record mode, which allows the trainer user to record their stroke for the trainee user to practice.
% (2) Practice mode, which allows the trainee user to select and practice the stroke.

% We created an AR interface to incorporate the customization for the visualization described in the previous section. 
% As shown in Figure \ref{fig:UI}(a), user initiates table tennis stroke training by selecting the pre-record stroke stored in the system database. 
% After determining the target training stroke, the user can now start the training by following the movement of the virtual expert avatar. 
% The \nickname will then render both visual cues and feedback. 
% For both the expert and user avatar, the user could drag and change the scale of the avatar with gestures.
% During the training, as shown in the Figure \ref{fig:UI}(b), the user can:
% \begin{itemize}
%     % \item change viewpoint of both the expert avatar and the user avatar by drawing and scaling the avatar with the free hand.
%     \item modify the stroke play of the expert avatar, user can pause, continue, speed up or speed down. 
%     \item vary the feedback provided by the system, such as enable or disable the body and paddle feedback.
%     \item change the motion error tolerance, the user can either increase or decrease the value of the threshold $\xi_{\text{joint}}$ and $\xi_{\text{paddle}}$, or choose adaptive tolerance.
% \end{itemize}

\begin{figure}[h!]
  \centering
  \includegraphics[width=\linewidth]{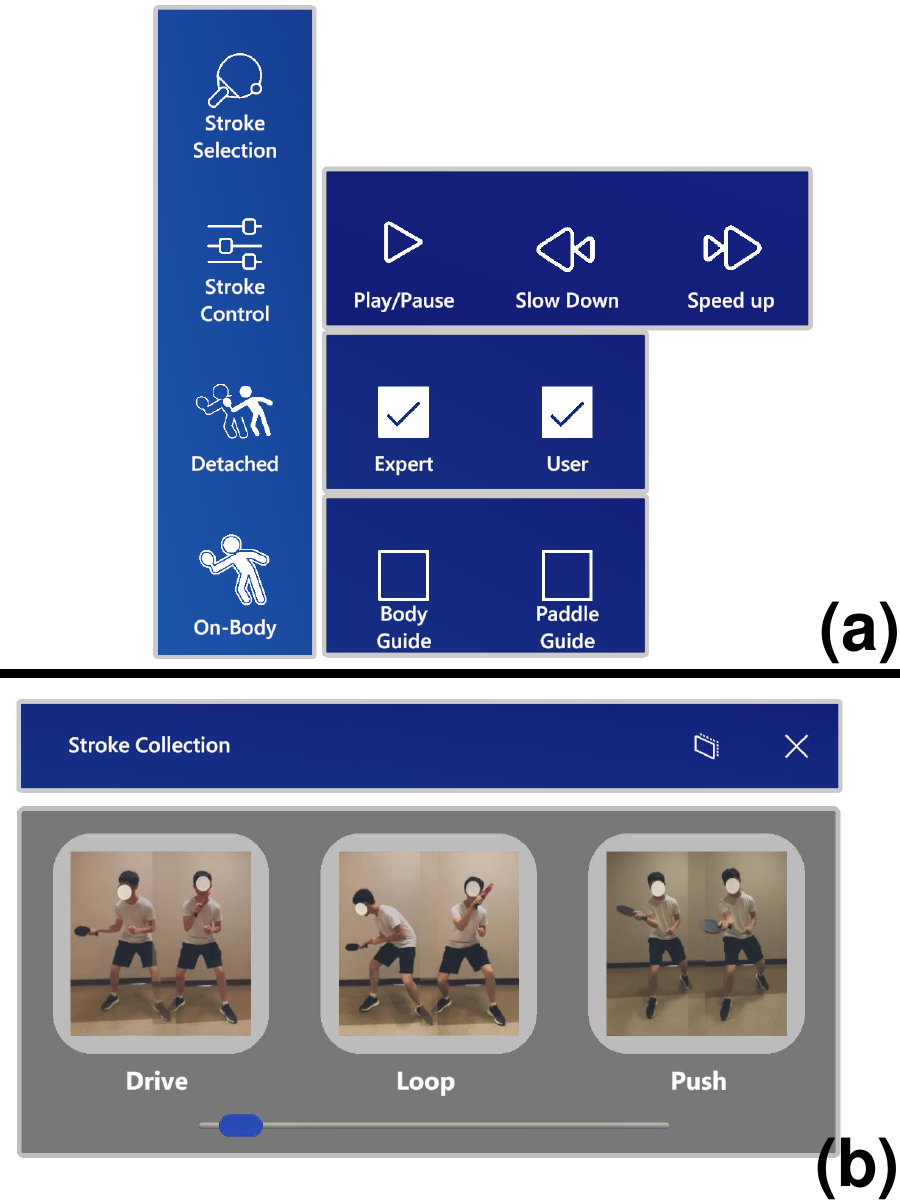}
  \caption{The AR interface of \nickname. 
  The user could adjusting their visualization preferences (a)
  and initiate the training by selecting the desired stroke from the menu (b) and  during the training session.}
  \label{fig:UI}
\end{figure}

We've designed an AR interface that offers customizable visualization options (C9), as outlined earlier. To begin practicing with \nickname, users simply click the "stroke selection" button located in the menu's first row, as seen in \autoref{fig:UI}(a). This action opens the stroke collection menu shown in \autoref{fig:UI}(b), where users can start their table tennis training by choosing a pre-recorded stroke from the system database.
Once a specific stroke is selected, users can observe and mimic the movements of the virtual expert avatar.
The \nickname~ system will provide both \detached~ and \onbody visuals based on the user preference.

For both the expert and user avatars, users have the capability to adjust the avatar's scale through gestures (C8). During the training, as depicted in ~\autoref{fig:UI}(a), users can:

\begin{itemize}
    % \item change the viewpoint of both the expert avatar and the user avatar by drawing and scaling the avatar with a free hand.
    \item modify the stroke play of the expert avatar; users can pause, resume, increase, or decrease the speed of the training.
    \item choose the \detached~ elements provided by the system, such as enabling or disabling expert avatar and user avatar. 
    \item customizes the \onbody~ feedback provided by the system, such as enabling or disabling body and paddle feedback.
\end{itemize}

%\

\section{Implementation details}
We implemented the \nickname~ system with a hardware setup consisting of a Microsoft Hololens 2 head-mounted display \cite{microsoft_2019_microsoft}, a 1080p HD Logitech webcam, an IMU equipped with a high-precision 9-axis gyroscope and Bluetooth 5.0 connectivity, along with a local PC equipped with an NVIDIA 1080Ti graphics card. 
An expert in the formative interview process was invited to record the motion of three stroke movements and the key static poses of each of the strokes, which are used to compare with the user movement and generate the expert avatar movement and \onbody~ cues.
For the user's movement, the visual input captured by the webcam is processed on the local PC with the 3D pose estimation algorithm \cite{mehta2017vnect}.
After receiving each frame from the camera, a request will be sent to the IMU (sample rate 250 HZ) to obtain the paddle pose. 
% Since the sample rate (250HZ) of the IMU  exceeds the pose estimation process by an order of magnitude, we make a simplified assumption that the paddle pose of the IMU corresponds directly to the estimated body pose from the camera for each frame.
Motion analysis is then applied to compare the user's stroke with that of the expert. The entire motion capture, analysis process, and visual cue rendering (the process that includes the three blocks to the left in ~\autoref{fig:SystemWorkflow}) are completed in approximately 36 ms (around 28 FPS) for each frame of the \detached~ user avatar and 17 ms (around 60 FPS) for the \detached~ expert avatar and \onbody~ cues which are faster because pose estimation is not needed.
The resulting visualizations are then rendered and presented through the HMD, utilizing Holographic Remoting with a connection latency of around 55 ms, for an immersive user experience. 
The \onbody~ cues are attached to the user based on their starting position, and the scales of the \onbody~ cues are adapted based on the height of the user; meanwhile, \detached~ cues are put 3 meters in front of the user, and the location and the scales of the \detached~ cues are changeable during use, allowing flexibility in visual representation.
The total latency of the system is around 91 ms, which is less than the visual reaction time of a table tennis player (around 260 ms \cite{bhabhor2013short, hulsdunker2019speed}).
% and tolerable game latency (around 100 ms) \cite{long2018characterizing}.
Both the authoring system and the learning system's user interface are constructed using Unity 3D, with some code components derived from \cite{3dUnity2023}. We empirically set $\xi_{joint}=\xi_{paddle}=0.1$ during system initialization.

\section{User Study}
We conducted a two-session user study to evaluate user learning outcomes and usability of the system. 14 users (12 identified as male and 2 identified as female, 21 to 28 years old) were recruited.
Of the 14 participants, 12 were familiar with VR applications on smartphones, tablets, or head-mounted devices, while the remaining two had prior exposure to both AR and VR technologies. 13 of the 14 participants did not have prior table tennis training experience, while one participant had received a moderate level of table tennis training before.  
The entire study lasted approximately 1.5 hours per participant, and each participant received compensation in the form of a \$15 e-gift card. Before diving into the study, participants were asked to get acquainted with Hololens2's interaction modality through its built-in tutorial.  
There are two sessions in our study. After each session, users completed a \added{5-point} Likert-type questionnaire \added{(Strongly Disagree; Slightly Disagree; Neutral; Slightly Agree; Strongly Agree)} about the training experience. At the end of the user study, each participant was interviewed and completed the standard System Usability Scale (SUS) questionnaire.

\subsection{Procedure}
% \begin{figure}[h]
%   \centering
%   \includegraphics[width=\linewidth]{figures/Figure_Procedure.pdf}
%   \caption{Ball return comparison, between the baseline approach and our AR method.}
%   \label{fig:BallReturn}
% \end{figure}

\begin{figure}[h]
  \centering
  \includegraphics[width=\linewidth]{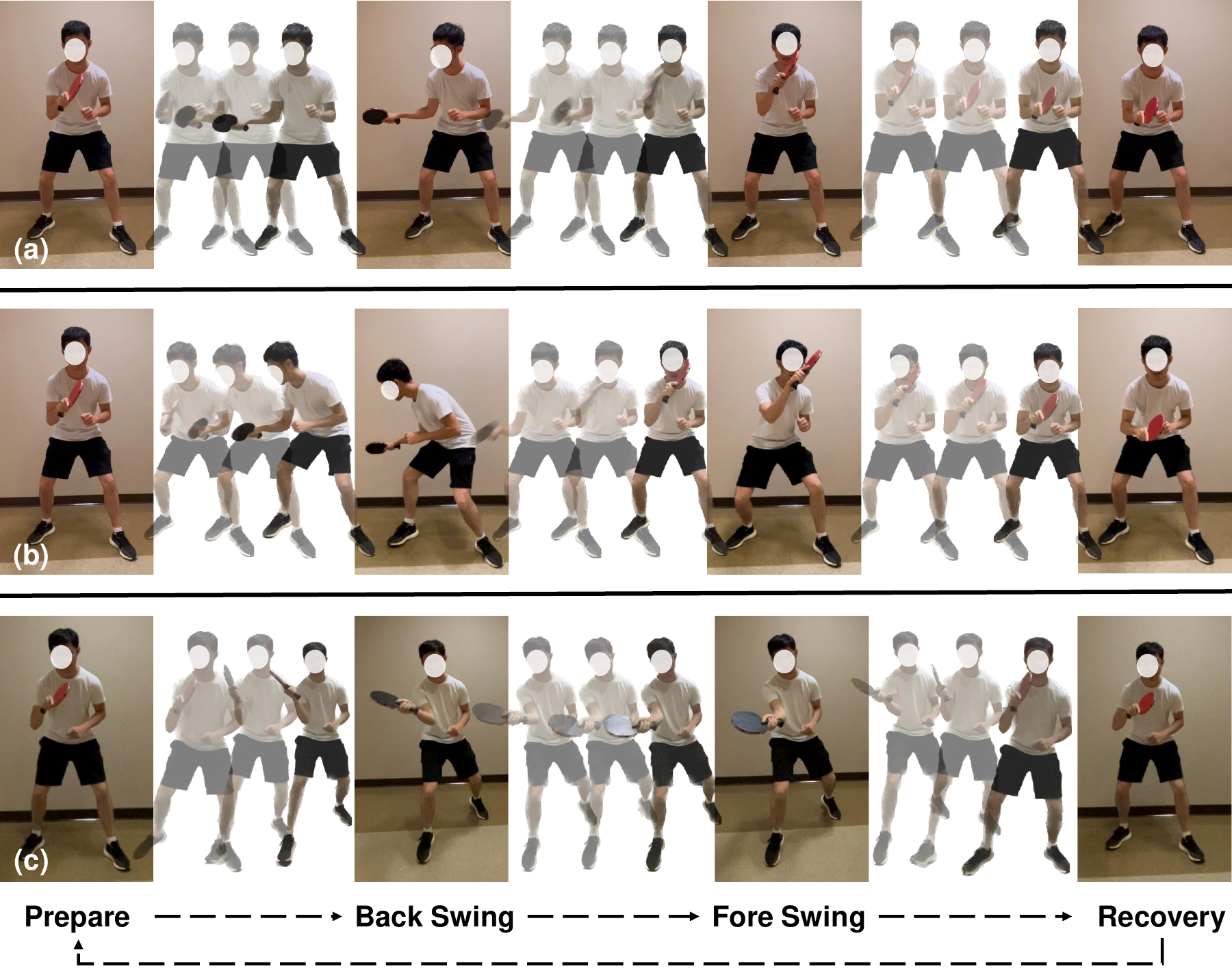}
  \caption{Forehand drive stroke (a), loop stroke (b), and push stroke (c) in the ``prepare - back swing - fore swing-recovery'' cycle (as mentioned in \textbf{I1}). We assess the entire movement and key postures, focusing on the prepare, back swing, and fore swing postures since the recovery posture matches the prepare posture.}
  \Description{A photo taken in Co-Rec}
  \label{fig:UserStudySetup}
\end{figure}

\subsubsection{Session 1: Movements Accuracy}
In this session, we aimed to evaluate the improvement in movement accuracy when using \nickname~ compared to a baseline system. Session 1 revolved around two similar strokes, namely drive and loop, structured as a between-subjects and counterbalancing experiment. This means that each participant was exposed to only one of the systems to learn a specific stroke. This approach ensured that the training results were not biased by the order in which the systems were used.
The two strokes were further decomposed into three key static poses of the stroke--preparation, back swing, and forward swing, as shown in ~\autoref{fig:UserStudySetup}. The user was asked to learn the three static poses first and then the whole stroke sequence. The total time spent with each system was the same.
For the experiment:
\begin{itemize}
    \item \textbf{Baseline System}: Participants observed an expert demonstration video. Subsequently, they attempted to adjust their posture and movement to mirror that of the expert. While doing this, they had access to a monitor displaying both their own video and the expert's, alongside detected skeletons. To facilitate learning, participants were allowed to adjust the camera angle and the monitor position by asking us. After a 5-minute practice session for each pose or sequence, their post-practice posture and movement were recorded. For each static posture, the demonstration time is 1 minute and the practice time is 3 minutes.
    \item \textbf{\nickname}: The participants began by observing the expert's demonstration video and the \detached~ skeleton. This setup allowed them to walk around the skeleton, observing posture and movement from various angles. The learning phase followed, in which the participants initially watched the \detached~ skeletons for 2 minutes, then the \onbody~ skeletons for another 2 minutes. At the last minute, they were given the freedom to choose which skeleton to observe. After this learning phase, participants were asked to perform poses or sequence three times. We then recorded their skeletons for analysis. For each static posture, the demonstration time is 1 minute and the practice time is 1 minute for \detached~, 1 minute for \onbody~, and 1 minute for free observation.
\end{itemize}
% Our analysis involved comparing the movements of participants exposed to both systems against expert skeletons using the methodology detailed in Section \ref{section:stroke_analysis}. 

\subsubsection{Session 2: User Experience}
Traditionally, learning a specific stroke in table tennis involves two primary steps:
\textbf{Observation and Shadow Practice}: Trainees observe a coach executing the stroke and then practice it without a ball, mimicking the movements.
\textbf{Multi-Ball Practice}: Trainees practice the stroke by returning multiple balls fed by a coach or a ball machine.
These steps are repeated iteratively during the stroke training process. 

In this session, our focus was on evaluating the user experience and overall usability of the \nickname~ system in the training cycle specifically for the first step, while utilizing a ball machine for multi-ball practice.

Our focus was on teaching the participants the ``push'' stroke with our platform. The experiment began with participants watching a video that briefly demonstrated the ``push'' stroke. After familiarizing themselves with the stroke through the video, they attempted to return balls launched from a ball machine. 
After an initial trial, participants engaged with the \nickname~ system. They observed and replicated the stroke using both \detached~ and \onbody~ cues, starting at a 50\% play speed. The speed gradually increased to full speed, and the participants practiced over 50 repetitions, spanning approximately 4 minutes. Subsequently, an additional minute was allotted for participants to freely interact with and use the system according to their preferences.

After using our system, participants once again tried to return balls from the ball machine, then filled out the system usability questionnaire and were interviewed.

\subsection{Results}

\subsubsection{Session 1 Results}
\label{para:s1results}
% motion improvement
\begin{figure}[h!]
  \centering
  \includegraphics[width=\linewidth]{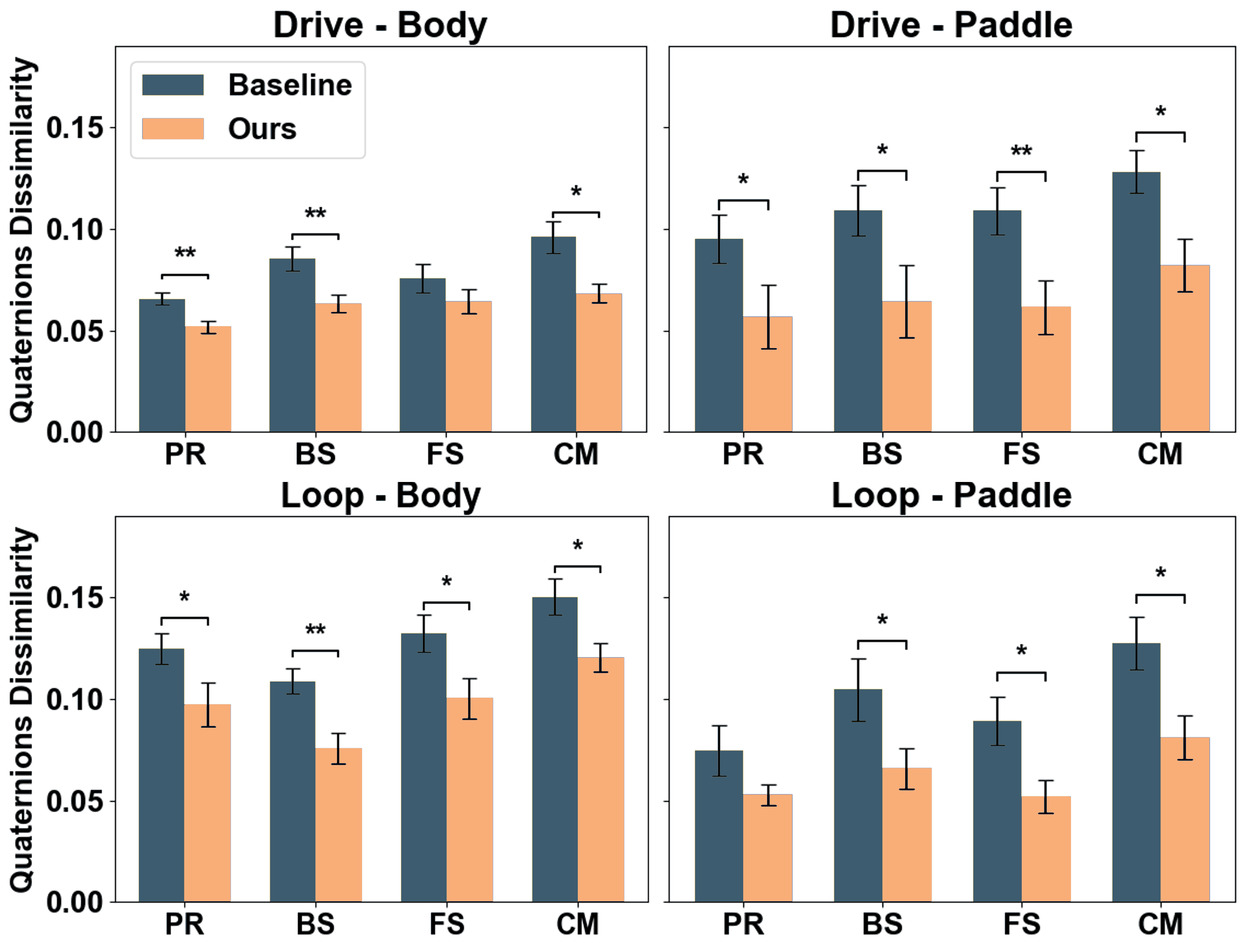}
  \caption{``prepare (PR) - back swing (BS) - fore swing (FS) - complete movement (CM)'' error comparison, between the baseline approach and our method, ``*'' indicates p-value < 0.05, ** indicates p-value < 0.01.}
  \label{fig:MovementCompare_individual}
\end{figure}

The outcomes of this session are presented in \autoref{fig:MovementCompare_individual}. Overall, user performance when performing static poses and movements was superior with \nickname~ compared to the baseline system. Specifically, for static poses, both body and paddle errors were reduced. This observation aligns with expectations, considering that a stroke sequence inherently incorporates multiple static poses. The average quaternion error for the baseline was 0.0987, while for \nickname~, it was lower at 0.0754. After discovering that the collected data are not normally distributed by performing a \textit{Shapiro-Wilk} test, a \textit{Wilcoxon signed-rank} test revealed substantial differences in pose errors between the two systems. However, exceptions were noted in the drive stroke ``FS'' body error and the loop stroke ``PR'' paddle error. The results suggest that \nickname~ offers a clear advantage in facilitating users to accurately replicate static poses. The incorporation of embodied and detached avatars in \nickname~ appears to enhance users' ability to mimic poses with greater precision compared to the baseline system.

\begin{figure}[h!]
  \centering
  \includegraphics[width=\linewidth]{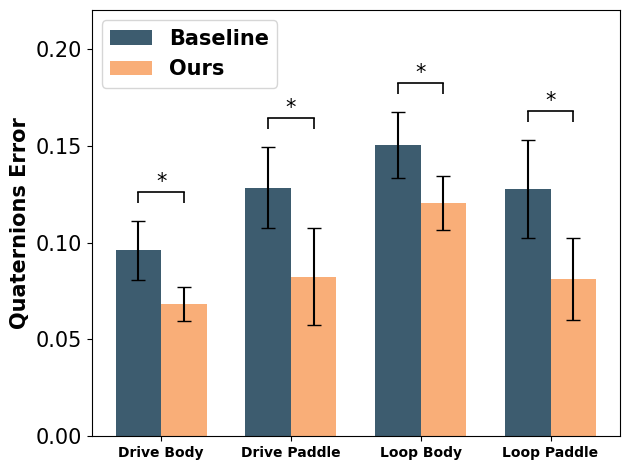}
  \caption{Overall the body and paddle movement error comparison (lower the better), between the baseline approach and our method, ``*'' indicates p-value < 0.05}%, ** indicates p-value < 0.01.}
  \label{fig:MovementCompare}
\end{figure}

When evaluating entire movements, the data in \autoref{fig:MovementCompare} further supports \nickname~'s efficacy. Users practicing with \nickname~ recorded a lower average error 
(drive stroke: body AVG = 0.0684, SD = 0.0249; paddle AVG = 0.0824, SD = 0.0709. loop stroke: body AVG = 0.1205, SD = 0.0391; paddle AVG = 0.0810, SD 0.0599)) 
in contrast to those who used the baseline system 
(drive stroke: body AVG = 0.0960, SD = 0.0427; paddle AVG = 0.1283, SD = 0.0587. loop stroke: body AVG = 0.1504, SD = 0.0479; paddle AVG = 0.1275, SD = 0.0712). 

\begin{figure}[h!]
  \centering
  \includegraphics[width=\linewidth]{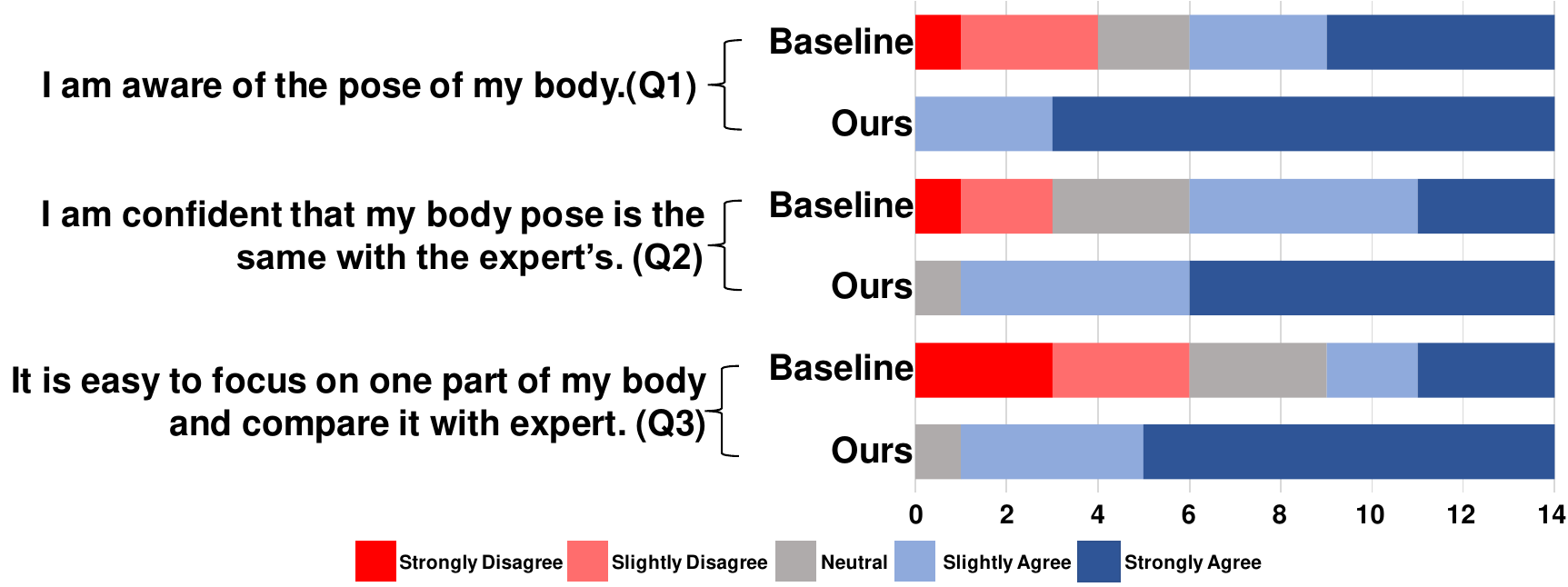}
  \caption{The results of the qualitative comparison between the baseline approach and our AR method.}
  \label{fig:Questionnaire_1}
\end{figure}

When comparing body and paddle errors for both strokes, we notice a more noticeable reduction in paddle error. This could be due to the \onbody~ skeleton that helps users better understand the movement of the target paddle. This observation aligns with \autoref{fig:Questionnaire_1}, which shows that users utilizing \nickname~ generally have an enhanced self-awareness of their body (Q1), increased confidence in the accuracy of their motion (Q2), and find it easier to compare their movements with those of the expert (Q3).
% usability

\subsubsection{Session 2 Results}
\label{para:s2results}

\begin{figure*}[h!]
  \centering
  \includegraphics[width=\linewidth]{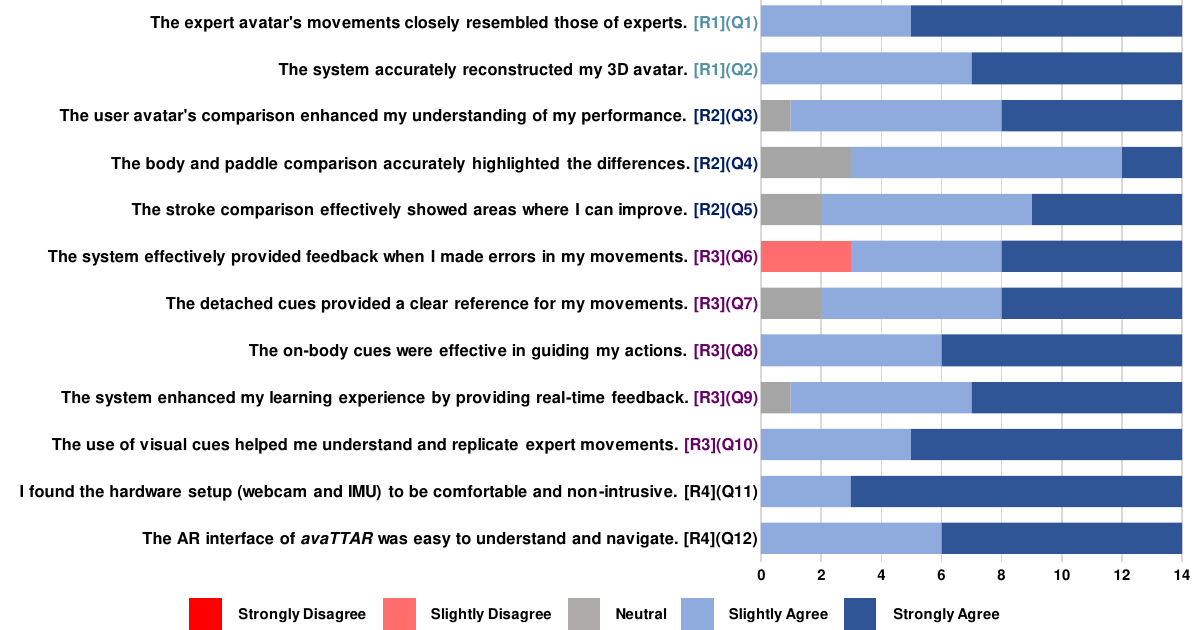}
  \caption{The results of the qualitative feedback on the system usability.}
  \label{fig:Questionnaire_2}
\end{figure*}

All 14 users successfully completed the training for the "push" stroke using \nickname. 

The features of the system were evaluated using Likert-type ratings collected at the end of this session, as shown in \autoref{fig:Questionnaire_2}. The questionnaire was divided into four parts, each part aiming to gather subjective user opinions on how \nickname~fulfills the design requirements described in Section ~\ref{DesignRequirements}.

Overall, users found our system to be highly user-friendly and navigable, with an average rating of 4.57 (SD = 0.51) for their ability to navigate through the system and manipulate avatars (Q12). Additionally, users reported that the hardware setup, including the webcam and IMU, did not significantly interfere with the training process, receiving an average rating of 4.79 (SD = 0.43) (Q11). A user commented, \textit{``It is intuitive to select the strokes and start using the system to learn the movement of the stroke.''}
Regarding the motion reconstruction features of our system, users expressed satisfaction with the representation of the expert avatar's stroke motion (Q1: AVG = 4.64, SD = 0.50) and their own avatar representing their body (Q2: AVG = 4.50, SD = 0.52). One user remarked, \textit{``The user avatar could follow my movement with low latency.''}
For stroke comparison, users' feedback was positive, indicating that the system effectively enhanced their understanding of their 
{stroke movement} 
performance (Q3: AVG = 4.36, SD = 0.63). According to one user, \textit{``Both the \detached~ avatars and the \onbody~ feedback are useful, showing me what should do and how to correct.''}
Users also found the comparisons between their body and paddle movements with those of experts to be accurate and informative (Q4: AVG = 3.93, SD = 0.62) (Q5: AVG = 4.21, SD = 0.70). As mentioned by a user, \textit{``I did some random movement that is not part of a stroke, the system immediately pointed out my error, that's fun.''}
Additionally, the system was reported to provide effective feedback when users made errors in their movements (Q6: AVG = 4.00, SD = 1.18), and the \detached~ cues provided clear references for user movements (Q7: AVG = 4.29, SD = 0.73). One user pointed out, \textit{``My avatar really highlighted where I was doing wrong during the stroke.''}
Users felt that the \onbody~ cues were helpful in guiding their actions (Q8: AVG = 4.57, SD = 0.51), and real-time feedback enhanced their learning experience (Q9: AVG = 4.43, SD = 0.65). One user observed, \textit{``The \onbody~ virtual paddle helped me understand the stroke well.''}
Users also found visual cues valuable in understanding and replicating expert movements (Q10: AVG = 4.64, SD = 0.50).
The standard SUS survey result is 80.89 with a standard deviation of 13.47, indicating the high usability of our system.

\section{Discussion and Future Work}

% \subsection{Embodied Sports Tutorial Authoring}

\subsection{Cues for Table Tennis Training}

The scope of our system is limited to the stroke training portion of table tennis, where accurate reproduction of stroke movement by the learner without guidance is of great importance. The result of our study (Section ~\ref{para:s1results}) indicates that the learners can accurately reproduce the stroke motion by using the system. 
Traditional practice can lead to reinforcement of incorrect techniques when practicing alone without accurate feedback, whereas \nickname~, with its demonstrations and guidance, ensures that learners practice movement posture correctly during training. 

\added{
Different levels of players can achieve different goals during stroke training. For beginner players, stroke training could help them develop and understand stroke techniques. For intermediate players, stroke training could help them reinforce their skills and adapt variants~\cite{artyomutochkin_2018_modern} of the same stroke to return incoming balls under different conditions such as different spins. For example, for the forehand drive stroke, the paddle angle is different with different kinds of spins. Another form of training, drills training, is designed to simulate match conditions and develop a combination of strokes (e.g. one forehand drive and one backhand drive). Our system also allows the user to record and train using these kinds of stroke variants and drills. 
}

% Moreover, \nickname~ provides a closer look at the stroke from the expert's viewpoint with the on-body cues. 
% This gives a similar experience of hands-on instruction that some of the coaches do when correcting the trainees.
% The significant improvement in error rates and performance metrics post-training with \nickname~—as opposed to those who only had repetitive practice —underscores the system's effectiveness in fostering true skill acquisition, not just familiarity through repetition.

As training often presents consistent and stable strokes for practice, there are potential gaps between the static way of table tennis training and the dynamic nature of real-world gameplay situations. 
Trainees may face challenges when trying to transfer the skills acquired during practice to actual gameplay scenarios. This is intrinsic to various training methods \cite{wu2021spinpong, oagaz2021performance} that exclude practical training with real opponents, including our system. 
As one user noted, \textit{``... the actual strokes are not always the same in real plays...''}. 
% Furthermore, the changes in return ball direction and landing position can lead to subtle differences in stroke execution. 
% A user expressed this concern \textit{``... the stroke could be different depending on where the ball is landed ...''}. 

Future work could involve adapting static target movements to different situations, such as changing ball trajectories and ball placements. Researchers could consider using generative models, such as GANs (Generative Adversarial Networks) \cite{goodfellow2020generative, goodfellow2014generative, shi2023hci} and diffusion models \cite{ho2020denoising, song2020denoising}, to enable such adaptations by synthesizing variant movements from static movements for the user to mimic. This dynamic training content could potentially adapt users to the situation in actual games.

Our system focuses only on the stroke training aspect of table tennis. In other aspects of table tennis training, more cues are needed in addition to body posture. Players could be interested in precise feedback on the contact points when the paddle hits the ball. A user with said: \textit{``... I want to know where I hit the ball if it was within the `sweet spot'...''}. Such cues provide detailed feedback for fine-tuning strokes. Additionally, timing cues are valuable to users, as mentioned by another user who expressed the need for \textit{``... an incoming virtual ball to help learn the timing to hit the ball...''}. These cues play a vital role in refining the timing and coordination required for advanced gameplay, improving the training experience for skilled players, which could lead to better performance or faster skill acquisition. 

% Future work in this area could involve the exploration of players at different skill levels and provide various aspects of training analysis, such as hitting performance (e.g., statistical analysis of ball-paddle contact area and hit timing). Additionally, incorporating multi-modality cues \cite{sigrist2013augmented, knoerlein2007visuo, wu2021spinpong} could further enhance the guidance provided to users. For example, haptic cues when the user is doing correctly and acoustic feedback indicating how to improve the pose.
% Furthermore, strategies related to ball placement could be integrated into the augmented reality experience. These advancements could be considered as an extension of previous work in the field of AR table \cite{ishii1999pingpongplus, Mayer_2019}, further enriching the training environment for players.
% As an extension of the current work, the integration of human expert evaluation could offer an understanding of learners' progress and ensure that the training aligns effectively with skill development.

To fulfill these requirements, we envision future work to consider outcome investigation, such as statistical analysis of ball-paddle contact area and hit timing. 
Furthermore, as an extension of previous work in the field of AR table \cite{ishii1999pingpongplus, Mayer_2019}, visualizing the trajectory and placement of the ball could be integrated into the HMD-based AR to enhance the training experience.
Additionally, multimodal cues could be considered to integrate to \nickname~ to improve user experience, such as haptic~\cite{wu2021spinpong, knoerlein2007visuo} and acoustic~\cite{sigrist2013augmented} feedback. 

\subsection{\Onbody~ Visuals for Other Sports and Areas}

The motivation of the \onbody~ visual is to provide a closer look at the stroke from the coach's viewpoint (first-person view of the coach) with the \onbody~ cues. 
This setup offers a similar hands-on instruction experience that some coaches have when correcting trainees, as one of the users \textit{``... the first-person view guidance, just like a coach, hand-to-hand correcting my posture...''}. 
The equivocation caused by the viewpoint was mentioned in \cite{liu2022posecoach}, where they provide a part-based visualization with viewpoint suggestions.
When relying on \detached~ cues, users must first mentally translate movements to their own viewpoint, which could lead to ambiguity.
The \onbody~ cues, on the other hand, illustrate the precise trajectories of both the body and the paddle to follow. 
A user said \textit{``... I found the \detached~ cues ambiguous sometimes, but \onbody~ ones are straightforward ...''}. 

% The \Onbody~ visuals that overlay on the physical body offer promising potential beyond table tennis. 
% Learning the body motions or hand-held object moving trajectory naturally has cognitive offset -- the drift of transfer objective motion from observation to execution. 
% The main characteristic of \onbody~ cues 
The \onbody~ visuals that overlay on the physical body may also be applied to other sports beyond table tennis. 
Especially those with basic skills that demand precise control of body movements. 
% For example, basketball involves a combination of such as dribbling, shooting, passing, etc. All of the fundamental skills require accurate motor functions and coordination.
For example, sports involve a combination of fundamental skills that are complex movements or fine motor skills: racket sports 
% \cite{ye2020shuttlespace, chu2021tivee} 
that have multiple elemental strokes similar to table tennis; basketball 
% \cite{lin2021towards} 
which involves basis movements such as dribbling, shooting, passing, and more.  
On the other hand, these visuals could be applied to train the rhythm in sports such as figure skating, where timing and rhythm are essential to execute jumps and spins with precision and control.

Similarly, we also notice that the \onbody~ cues for the paddle improve the spatial understanding of the object for the learner, as mentioned by a user, ``... I found the \onbody~ cues for the paddle alone are very useful to understand the orientation and path of it ...''
Training scenarios that involve the use of tools or hand-object interaction \cite{jain2023ubi} in various domains could also consider using \onbody~ instructions in future studies, such as welding \cite{ipsita2022towards}, carpentry \cite{lee2020using}, or surgery \cite{vavra2017recent}, could also benefit from a similar MR training approach, improving skill acquisition, safety, and precision in professions where effective tool usage is critical. 
% For general motor skill learning, \onbody~ visualization could also help understand subtle movements such as wrist movement and grips in various sports. 
% Investigating the transferability of our system across different sports and tool-handling scenarios would not only expand its application, but also contribute to a more versatile and adaptable training tool for users across various domains.

\subsection{Movement Reconstruction Limitation}
The motion reconstruction component of \nickname~ provides promising results overall, according to the user study. However, there were cases where the user's physical movements mismatched their virtual representation within the system. A user who wore clothes in a white color that is close to the color of the background mentioned that \textit{``... my avatar's leg is jittering when my leg is not...''}. Also, since the pose estimation network operates at a frequency of around 30 frames per second (FPS), there might be latency issues with rapid movement. As expressed by another user, \textit{``...my avatar cannot follow me when I move my hand fast...''}. These examples illustrate the potential challenges posed by pose estimation errors. While in the training phase, the latency and the low FPS might be ignored, they more or less interfere with the user experience.

Limited by the graphic card, \nickname~ adopted a lightweight neural network with acceptable performance. To obtain better reconstruction accuracy, future studies should consider using advanced algorithms \cite{li2022mhformer, zheng20213d, foo2023unified, gong2023diffpose}. However, these methods may require more computational resources and result in a low FPS. Future solutions may involve the use of cloud services to process data to solve resource challenges.

\added{
In table tennis, the grip remains unchanged during stroke execution, therefore, the hand posture is also fixed. 
The position and orientation of the paddle can also deduce the position and orientation of the wrist. 
Therefore, we did not incorporate the detection of the hand pose. 
However, this might limit the understanding of the wrist and hand movement.
Future studies could apply whole-body estimation~\cite{zhu2023h3wb} to detect and provide the visualization of both body and hand. 
}

\subsection{Field of View}

Intrinsic to the \onbody~ and \detached~ visualization method, only when the natural viewing area aligns with the movement that the user attempts to learn, visualization can be applied efficiently.
Otherwise, the user might not be able to view and practice the correct motion at the same time.
This viewing area problem has been investigated in~\cite{jo2023flowar}, where the researchers fixed the visualization to various positions relative to the user.

Besides the software limitation on the field of view, the current system may suffer from a limited field of view with HoloLens which may compromise the overall training experience. 
In particular, a comment arose with the \onbody~ cues, \textit{``... I can only see parts of the \onbody~ visual cues, such as my hand and paddle...''}. 
Future work could explore the use of novel AR/MR headsets, such as the Meta Quest Pro~\cite{meta} and Apple Vision Pro~\cite{apple}. These headsets offer a wider field of view, potentially improving the immersive quality of the system and addressing this limitation. 
\added{Additionally, the weight and comfort of the HMD would also be one of the possible reasons that hinder the performance of the user, which can also be investigated in the future.}

\added{
\subsection{Visual Presentation}
Based on the design rationale, we adopted the skeletal visualization by following YouMove~\cite{anderson2013youmove} which also includes learning body movement. 
According to \cite{yu2024design}, based on the level of indirection there are three types of motion guidance visualization which are explicit (e.g. YouMove~\cite{anderson2013youmove}), implicit (e.g. SleeveAR~\cite{sousa2016sleevear}), and abstract (e.g. LightGuide~\cite{sodhi2012lightguide}). 
Future work could consider studying alternative visualization methods based on these three categories.
For example, for players who already understand stroke movement, explicit skeletal motion guidance could be redundant. 
Using only the key joint movement path or racket movement, implicit (or abstract) visualization could also be applied. 
Additionally, for explicit visualization, it is unknown whether the skeletal method or the avatar method~\cite{ikeda2018ar} is more effective, researchers could consider exploring user preferences for body movement visualization.

Visual attention guidance techniques \cite{el2009dynamic} could also be beneficial when visualizing the stroke. Although our system did not employ advanced visual attention guidance methods, studies \cite{bautista2023strategies} have demonstrated that directing the user's attention to critical areas, such as errors or difficult segments of the stroke, can help reduce cognitive load.
}

\section{Conclusion}
 In this work, we presented \nickname, an AR system that provides \onbody~ and \detached~ visual cues for the training of table tennis strokes. This dual visualization approach not only strengthens the user's understanding of the correct techniques but also offers immediate feedback for refinement. Our contributions include a design rationale extracted from interviews with experienced players. Based on the design rationale, we derived the design requirements for our system and then decomposed them into detailed components to implement. Our system integrated a camera and IMU setup to capture 3D body and paddle movements, and an AR interface that enables users to practice with personalized visual cues. Furthermore, our user study first highlights the potential of \nickname~to improve movement posture accuracy, then the usability of the system to practice stroke movements. We envision that future research can further investigate other types of visual cues and apply the cues we propose to other areas.

\section{Acknowledgements}
We give thanks to the reviewers for their invaluable feedback. 
This work is partially supported by the NSF under the Future of Work at the Human Technology Frontier (FW-HTF) 1839971. 
We also acknowledge the Feddersen Distinguished Professorship Funds. 
Additionally, we would like to thank the coach and players from the Purdue Table Tennis Club and the Indy Pong Table Tennis Center for their invaluable support.

\bibliographystyle{ACM-Reference-Format}
% \bibliography{ref}

\appendix
\added{
\section{Formative Interview}
}
Questions asked in the interview:
\begin{enumerate}
    \item Do you have any specific stroke training experience in table tennis? If yes, please briefly describe the type of stroke training you have undergone and any notable outcomes.
    \item What other types of training or practice have you engaged in to improve your table tennis skills (e.g., footwork, serve and receive practice, tactics and strategy training, physical conditioning, ball control)?
    \item In your opinion, what are the critical components of an effective stroke training program?
    \item What specific aspects of table tennis stroke training do you find most challenging?
    \item What are your preferred methods or resources for learning new table tennis techniques? (Videos, books, in-person coaching, etc.)
    \item Have you ever had experience coaching or mentoring others in table tennis or received formal training or coaching from a table tennis coach?
    \item Briefly describe your coaching (from) others' experience (e.g., duration, level of players, specific areas of focus).
    \item (For coach) Based on your coaching experience, what are some common challenges you have encountered when coaching table tennis strokes? How do you currently address or overcome these challenges?
    \item (For trainee) How do you believe training with a coach has impacted your table tennis stroke technique and overall performance? Please provide specific examples or instances where you felt the coaching had a significant influence.
    \item What kind of feedback or guidance do you find most helpful when practicing table tennis strokes?
    \item Have you ever used any technology or applications to aid in your table tennis stroke training? If yes, please provide details.
    \item Have you ever used virtual reality (VR) or augmented reality (AR) technologies for sports training? Have you heard about VR or AR for sports training? If yes, please describe your experience and its impact on your training.
\end{enumerate}

\begin{table}
  \caption{Demographics of interviewees, including age, table tennis experience in years, training experience with a coach in years, and the insights they mentioned. Only P6 and P10 have AR / VR experience.}
  \label{table:tt_training_data}
  \begin{tabular}{ccccl}
    \toprule
    \textbf{ID} & \textbf{Age} & \textbf{TT (Years)} & \textbf{Training (Years)} & \textbf{Insights} \\
    \midrule
    P1  & 69 & 54 & 2  & I2, I3 \\
    P2  & 29 & 19 & 1  & I2, I3 \\
    P3  & 65 & 25 & 5  & I2, I3 \\
    P4  & 66 & 20 & 9  & I1, I2, I3 \\
    P5  & 47 & 35 & 10 & I1, I2, I3 \\
    P6  & 27 & 20 & 12 & I1, I2, I3 \\
    P7  & 26 & 20 & 4  & I1, I2, I3, I4 \\
    P8  & 27 & 8  & 0  & I2, I4 \\
    P9  & 23 & 6  & 0  & I2, I3, I4 \\
    P10 & 22 & 1  & 0  & I2, I3 \\
    P11 & 24 & 12 & 9  & I1, I2, I3, I4 \\
    \bottomrule
  \end{tabular}
\end{table}

\end{document}